# FLUORESCENT NANOPARTICLES FOR SENSING

*Anil Chandra[1], Saumya Prasad[1], Giuseppe Gigli[1,2] and Loretta L. del Mercato[1,*]*
1 CNR NANOTEC – Institute of Nanotechnology c/o Campus Ecotekne, via Monteroni, 73100, Lecce, Italy
2 Department of Mathematics and Physics "Ennio De Giorgi", University of Salento, via Arnesano, 73100, Lecce, Italy

*Corresponding author: loretta.delmercato@nanotec.cnr.it

## ABSTRACT

Nanoparticles-based fluorescent sensors have come up as a very competitive alternative to small molecule sensors due to their exquisite fluorescence-based sensing capabilities. The tailorability in design, architecture and photophysical properties has attracted the attention of many research groups which has resulted in many reports related to novel nanosensors applied for sensing a vast variety of biological analytes. Although semiconducting quantum dots have been symbolized as a strong representative of fluorescence nanoparticles for a long time, an increasing number of reports on new classes of organic nanoparticles-based sensors, such as carbon dots and polymeric nanoparticles, have shown their increasing popularity due to their biocompatibility, ease of synthesis and biofunctionalization capabilities. For instance, fluorescent gold and silver nanoclusters have come up as a less cytotoxic replacement of semiconducting quantum dots sensors. This chapter provides an overview of recent developments in nanoparticles-based sensors for chemical and biological sensing and includes a discussion on unique properties of nanoparticles of different composition along with their basic mechanism of fluorescence, route of synthesis, as well as their advantages and limitations.

## 1. INTRODUCTION

The phenomenon of fluorescence can be described in a simple way as the process of emission of light when a molecule absorbs light of a certain frequency. In most cases, the emitted light has a longer wavelength than the absorbed light. John Herschel reported fluorescence from quinine water for the first time in 1845.[1] As fluorescence emission is highly sensitive towards the immediate microenvironment of a fluorophore, it has been exploited in the development of several sensing applications. Any parameter which could interact with the fluorophore at the molecular level would affect its fluorescence.[2] Some critical chemical and physical parameters effecting fluorescence are shown in **Figure 1**. Several molecular probes (natural and chemically synthesized) have been reported to exhibit more sensitivity towards a few selected parameters. For example, an ideal fluorophore for pH sensing should exhibit its highest sensitivity towards pH changes and should be less responsive towards other microenvironment parameters. The growing number of publications in the field of fluorescence (**Figure 2**) emphasizes the importance of fluorescence in different fields of research. This chapter will mainly focus on fluorescent nanoparticles (NPs) (also denoted as nanosensors) for bio-diagnostic applications. Based on their composition, nanosensors could be categorized broadly into inorganic and organic nanosensors (**Figure 3**). A long list of fluorescent NPs exists in literature, but for simplicity we have included only a few selected categories in this chapter. For a more elaborate understanding of fluorescent NPs used in sensing, the readers can follow several exhaustive reviews dedicated to different classes of sensors.[12–15]

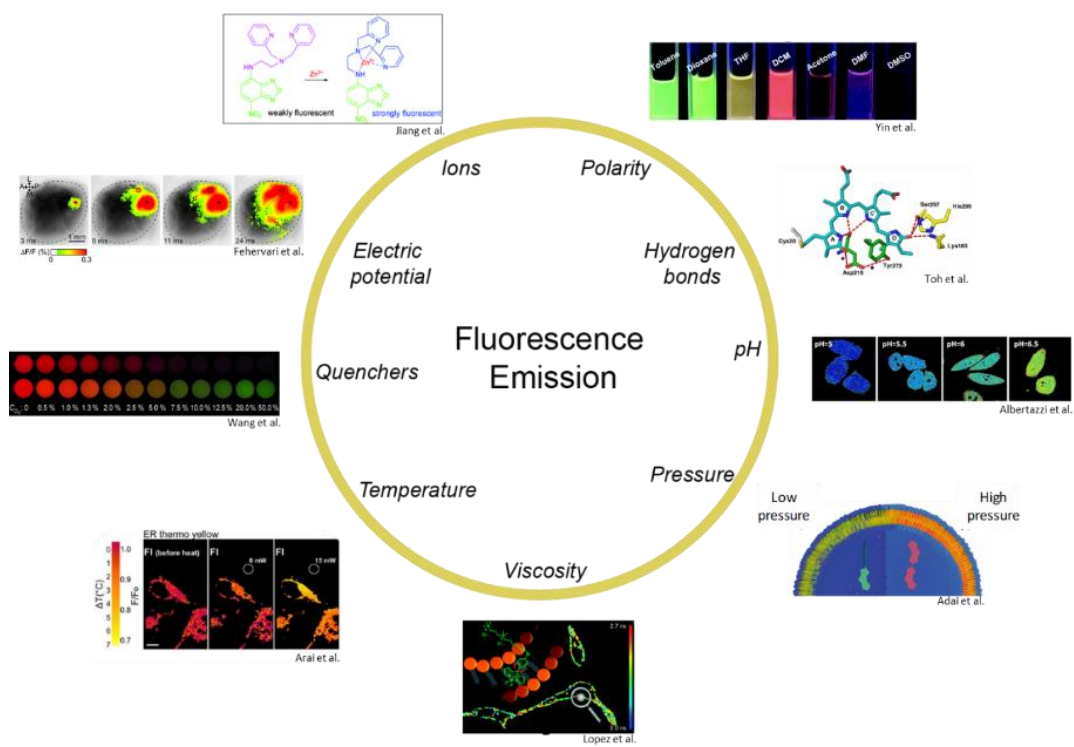

**Figure 1.** Different physical and chemical parameters that can influence fluorescence emission.[3–11]

A key aspect of fluorescence-based sensing is the size of the sensing system. It could be either a small molecule, nanoparticle, microparticle, film or fiber. To get an idea about the dimensions of the nanoprobes, different length scales with representative examples are shown in **Figure 4**. The size of the fluorescent nanosensors also becomes an important factor when sensing has to be done in a biological environment. A small molecule typically falls in the angstrom size range and are free to diffuse, whereas NPs are bigger objects (1-100 nm) and diffuse slowly. For example, enhanced permeability and retention effect (EPR) involves the selective accumulation of high molecular weight materials (e.g., NPs, micelles, liposomes) in the tumor that can help in tumor bioimaging and targeting.[16,17] NPs-based fluorescent sensors offer plethora of advantages compared to small molecule-based sensors. Their advantages include easy biofunctionalization, tunable fluorescence properties, multianalyte sensing capability and their ability of ultrasensitive detection at even nanomolar and picomolar particle concentration which is not possible using small molecules.[18,19]

The origin of fluorescence in a sensor could vary depending on their size range. For example, fluorescence from a NP could originate by virtue of its intrinsic fluorescence because of quantum confinement, as observed in semiconducting quantum dots,[20] or due to linking of small fluorescent molecules/dyes to it.[21] In another class of NPs which are organic (carbon-based NPs), the origin of fluorescence is still debatable and could be due to quantum confinement[22,23] or emissive domains[24] or both.[25] In the case of micron-sized particles, fluorescence arises as a result of other fluorescent molecules or NPs that are linked to the microparticles. [26–31]

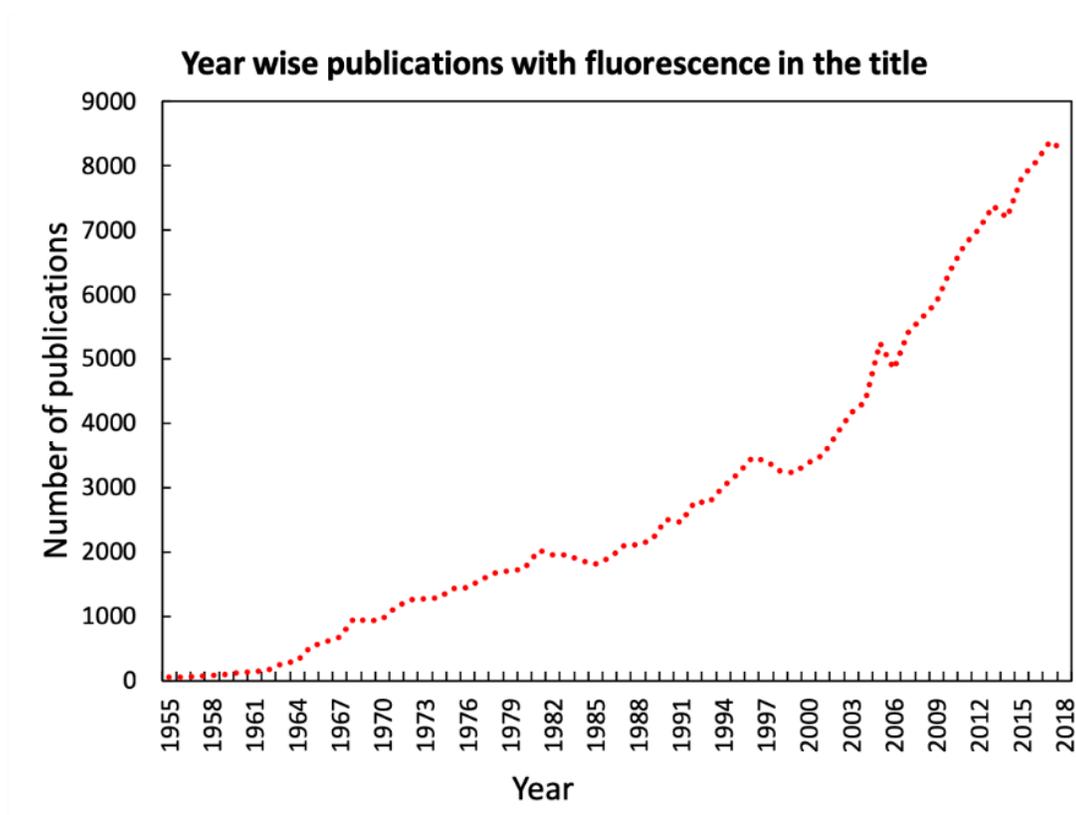

**Figure 2.** Year-wise fluorescence related publications between 1955 and 2018 (data collected from Web of Knowledge).

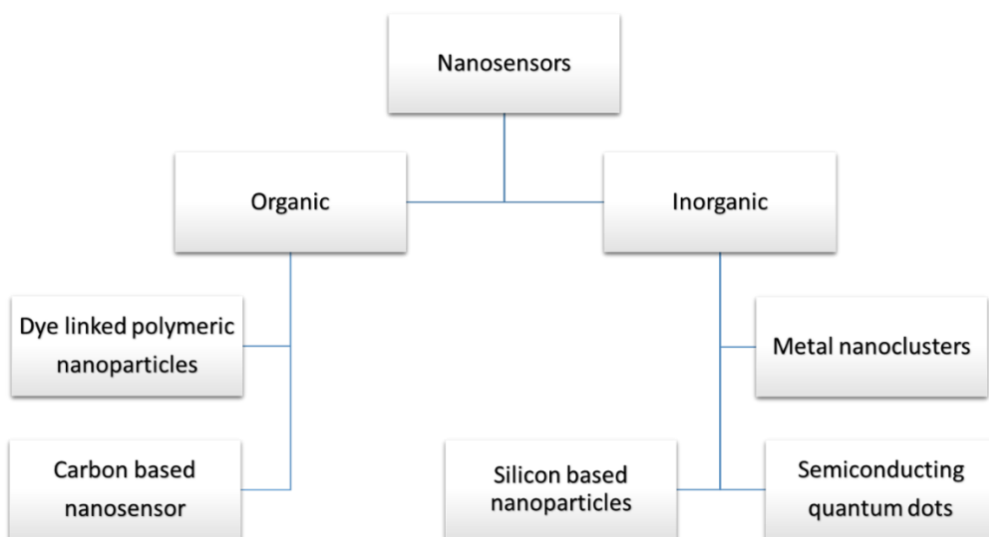

**Figure 3.** Sub-classification of particle-based sensors based on the composition and origin of fluorescence.

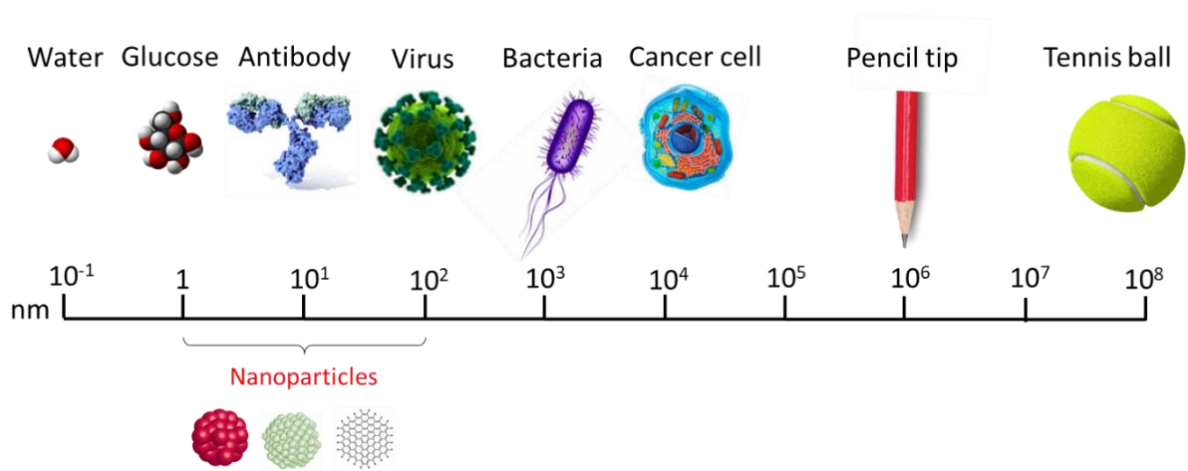

**Figure 4.** Relative size range of NPs compared with objects of different dimensions.

In the past, the application of fluorescence-based sensing and detection has been realized by using a variety of approaches which are based not only on the change in emission intensity as a read-out. Notably, a simple fluorescence emission intensity can be linked to plenty of other processes that can affect it. For example, a parameter of interest could be sensed by either observing the change in fluorescence emission due to quenching or due to a change in fluorescence lifetime. In some cases, change in the diffusion coefficient of a dye can also be used for sensing. Whatever the method is, the raw observation is a change in fluorescence emission properties and all the sensing techniques exploit one or the other ways to detect this change. A very common example of fluorescence-based sensing could be the use of fluorescein for sensing pH in a solution. The reason for fluorescein's variable emission intensity with change in pH is the presence of ionizable functional groups in its structure, due to which it exists in an ionic equilibrium consisting of different forms like dianion, monoanion, neutral, cation and lactone. These different forms have different fluorescence emission properties and thus variation in ionic equilibrium due to change in pH can sharply change the overall fluorescence of the solution.[32] Förster resonance energy transfer or fluorescence resonance energy transfer (FRET) based sensing is another approach that can be exploited for sensing applications. It involves energy transfer from an excited donor to an acceptor that can result in quenching of donor, which causes a decrease in its fluorescence emission intensity.[33–35] The acceptor, on the other hand, can use the energy collected from the donor to exhibit enhanced emission intensity or can remain nonfluorescent by releasing the energy by non-radiative relaxation.[36] This phenomena of FRET is extremely sensitive towards the distance between the donor and acceptor chromophores, thus it can be used in studies involving molecular interaction in chemistry[37] and biology.[38] For its use in assessing the distance between biomolecules, it is often called as a spectroscopic ruler.[39]

**Table 1** describes examples of different categories of NPs-based fluorescent sensors with information about their size, fluorescence excitation/emission and the analyte they can sense. They are discussed in more detail along with their mode of sensing in the later part of this chapter.

**Table 1.** Examples of fluorescent NPs for sensing bioanalytes.

| S.No. | Semiconducting Quantum Dots | Size | Fluorescence Excitation & Emission Maxima | Sensing Analyte |
|---|---|---|---|---|
| 1 | CdSe@SiO2@CdTe | 44.5 ± 5.0 nm | Ex: 380 nm; Em: 522 nm, 616 nm | Ascorbic acid[40] |
| 2 | L-cys ZnS:Mn QDs (L-cysteine capped Mn-doped ZnS QDs) | ~2.8 nm | Ex: 302 nm; Em: 598 nm | Dopamine[41] |
| 3 | BSA–AgNCs (Bovine serum albumin-confined Ag NCs) | 2 nm | Ex: 465 nm; Em: 598 nm | Biothiols[42] |
| 4 | ⁻OPhS-QDs (sodium 4-mercaptophenolate functionalized CdSe/ZnS QDs) | 4.1 ± 0.4 nm | Ex: 450 nm; Em: 582 | pH[43] |
| 5 | CdSe/ZnS core-shell QDs | 2-4.5 nm | Ex: 485; Em: 525 nm | Glucose[44] |
| | **Gold and Silver Nanocluster Sensors** | **Size** | **Fluorescence Excitation & Emission Maxima** | **Sensing Analyte** |
| 1 | AuNCs and Au@AgNCs | AuNCs (1.8 nm), Au@AgNCs (2.4 nm) | AuNCs (Ex: 365nm; Em: 670nm), (Au@AgNCs Ex: 365 nm; Em: 565 nm) | Cu(II)[45] |
| 2 | AuNCs | <1nm | Ex: 475 nm; Em: 640 nm | Cyanide in water[46] |
| 3 | Fib-Au NCs (fibril stabilized gold NCs) | 1.6 ±0.4 nm | Ex: 520 nm; Em: 675 nm | Cysteine[47] |
| 4 | (Lysozyme Type VI)-Stabilized Au 8 clusters | 13±1 nm | Ex: 380 nm; Em: 455 nm | Glutathione[48] |
| 5 | HRP-AuNCs | 2.7±0.6 nm | Ex: 365 nm; Em: 450nm, 650 nm | Hydrogen peroxide[49] |
| 6 | PMAA-AgNCs | < 1 nm | Ex: 510 nm; Em: 615 nm | Cu(II)[50] |
| 7 | Nuceic acid stabilized AgNCs (two types of AgNCs) | 5-6 nm | Ex:480 nm; Em: 560 nm, Ex: 520 nm, Em: 615 nm | Genes (Werner syndrome, HIV)[51] |
| | **Silicon based Nanosensors** | **Size** | **Fluorescence Excitation & Emission Maxima** | **Sensing Analyte** |
| 1 | probe–MSN–PEG | 120 nm | Ex: 370 nm; Em: 418 nm | Homocysteine[52] |
| 2 | Ratiometric fluorescent probe-doped silica NPs (PL1-SiO2) | 53-60 nm | Ex: 360 nm; Em: 470 nm Ex: 420 nm; Em: 540 nm | Hydrogen peroxide[53] |
| 3 | Dual-modified SiNPs (DMSiNPs) | ~8.8 nm | Ex: 405 nm; Em:465 nm Ex:543 nm; Em:575 nm | pH[54] |
| | **Carbon Nanosensors** | **Size** | **Fluorescence Excitation & Emission Maxima** | **Sensing Analyte** |
| 1 | CDs (derived from citric acid and basic fuchsin) | 5-12 nm | Ex: 380 nm; Em: 475 nm, 545 nm | pH[55] |
| 2 | GQDs | Diameter 2.8 nm, height 0.5-1 nm | Ex: 460 nm; Em: 527 nm | Urea[56] |
| 3 | N-doped CDs | ~5–7 nm | Ex: 405 nm; Em: 450–550 nm | ALP Activity[57] |
| 4 | Boronic acid modified CDs | 2.5 nm to 6.5 nm | Ex: 320 nm; Em: 408 nm | Glucose sensing[58] |
| 5 | Nnaphthalimide-azide linked CDs | ~5 nm | Ex: 340 nm; Em: 425 nm, 526 nm | Hydrogen sulfide[59] |
| 6 | N-doped CDs | 5.18±1.3 nm | Ex: 365 nm; Em: 450 nm Ex: 500 nm; Em 550 nm | Intracellular pH[60] |
| | **Polymeric Nanosensors** | **Size** | **Fluorescence Excitation & Emission Maxima** | **Sensing Analyte** |
| 1 | Cyclen-functionalized fluorescent polymeric NPs | 33 nm to 40 nm | Ex: 490nm; Em: 540 nm | Copper ion and sulfide anion[61] |
| 2 | Pluronic based organic nanoparticle-NIR luminogen | 109 nm | Ex: 550 nm; Em: 654 nm | Cysteine[62] |
| 3 | DNA-Functionalized dye-loaded polymeric NPs | ~40 nm | Ex: 530 nm; Em: 580, 665 nm | Nucleic acids[63] |
| 4 | Poly (DOPA) fluorescent pH-sensor | ~27 nm | Ex: 310 nm; Em: 410, 500 nm | pH[64] |

## 2. INORGANIC NANOSENSORS

### 2.1 Semiconducting quantum dots

Quantum dots (QDs) are semiconducting tiny crystals with a diameter ranging from 2-10 nm. They are highly fluorescent and have been employed for various bioimaging and sensing applications.[65–67] QDs are advantageous due to their exquisite fluorescence properties which are due to quantum confinement effect.[68] They can be excited using a broad range of wavelengths which provides flexibility in terms of instrumentation and light source.[20,65] On the other hand, their emission is very defined within a very specific and narrow bandwidth which makes them very useful for multiplexed sensing with minimal overlap in emission signals. Another distinguishing feature of QDs is their exceptionally high quantum yields (QY) due to which they can be used in very small number to generate a detectable fluorescence signal. In general, semiconducting QDs are synthesized by the solvothermal approach, where precursors are crystallized in coordinating solvents at high temperature. The size of the QDs is controlled by adjusting the number of precursors and crystal growth time. The fluorescence properties of semiconducting QDs is highly dependent on their surface, thus epitaxially passivating the surface of QDs with higher bandwidth species like zinc sulphide (ZnS) has been observed to enhance the QY from 10% to 80%.[69,70] The passivating layer also protects the QDs from degradation which can release toxic metal ions into the surrounding media.[71]

Semiconducting QDs have been utilized profoundly for the development of fluorescence-based sensing and bioimaging platforms. Since the first application of cadmium selenide (CdSe) QDs for F-actin labeling in 1998,[20] plenty of other applications have been explored overtime for labeling and/or sensing biomolecules both *in vitro* and *in vivo*. The synthesis of inorganic QDs in organic solvents involves the use of hydrophobic stabilizing molecules, which coats their surface. Thus, biological application of these QDs requires them to be hydrophilic. As a solution, Pellegrino *et al.* developed a simple and general strategy for decorating hydrophobic nanocrystals of various materials ($CoPt_3$, Au, CdSe/ZnS, and $Fe_2O_3$) with a hydrophilic polymer shell by exploiting the nonspecific hydrophobic interactions between the alkyl chains of poly(maleic anhydride alt-1-tetradecene) and the nanocrystal surfactant molecules.[72] As another example of enhancing the hydrophilicity, Argüelles *et al.* reported polymer-coated QDs with integrated acceptor dyes that can act as FRET based nanoprobe.[73] Their study showed the importance of polymer coating not only for colloidal stability but to enhance the sensitivity of the probe. In their work, they stressed upon the efficient working of QD based FRET probes by changing the general arrangement of a QD based FRET sensor, which in general consist of a QD donor with acceptor molecules linked to it via a receptor ligand interaction. The sensitivity in this kind of FRET probes is conferred by changing the distance between the donor and the acceptor in presence of a certain analyte that in turn changes the FRET efficiency. In this work, they introduced a FRET geometry in which the acceptor (ATTO590 dye) is directly incorporated into the encapsulation shell used to provide colloidal stability to the donor (Qdot 545 ITK). This caused a very stable linkage between the donor and the acceptor. Their approach had many advantages like efficient acceptor binding, the binding possibility for both hydrophilic as well as hydrophobic acceptor molecules and excellent colloidal stability without compromising the FRET efficiency.

Glucose sensing has always been essential for the diagnosis of various metabolic disorders. Although there are reports on spectrophotometric[74] and fluorometric based glucose sensing methods,[75,76] Duong *et al.* for the first time created a QD based glucose sensor. They took advantage of the enzymes, glucose oxidase and horseradish peroxidase and created an enzyme conjugated QD which was used as a FRET-based glucose sensor.[44] The transfer of non-radiative energy from the QDs to the enzymes resulted in the fluorescence quenching of the QDs, corresponding to an increase in the concentration of glucose. The linear detection ranges of glucose concentrations were 0–5.0 g L$^{-1}$. The probe required 30 minutes to reach stability which was slower compared to membrane based glucose sensors developed by other authors in their previous studies with a response time of 0.15–0.5 min.[77]

Recently Xu *et al.* reported pH sensitive sodium 4-mercaptophenolate capped CdSe/ZnS QDs (denoted as $^-$OPhS-QDs).[43] $^-$OPhS-QDs exhibit strong fluorescence in near neutral medium, whereas $^-$OPhS- moieties on QDs surface easily binds to proton under acidic conditions to yield 4-mercaptophenol capped QDs (i.e. HOPhS-QDs), which acts as an efficient hole trapper. As a result, the photoluminescence (PL) of QDs is switched off. The pH probe exhibits a good linear relationship between fluorescence response and pH values in the pH range of 3.0–5.2. This sensing platform has potential of sensing in highly acidic conditions that exist in tumors due to the Warburg effect.[78]

In another work, ascorbic acid (AA) sensing using a dually emitting CdSe@SiO$_2$@CdTe QD hybrid was reported by Wang and colleagues. The nanohybrid consists of red-emitting CdTe QDs covalently linked to the surface of silica NPs containing green-emitting CdSe QDs. Two types of QDs were used, where the green-emitting CdSe QDs were used as a reference and red CdTe QDs as AA sensors. The fluorescence of red QDs is first quenched using KMnO$_4$, which causes oxidation and formation of CdTeO$_3$ and TeO$_2$. On addition of AA, the red fluorescence is restored due to the conversion of CdTeO$_3$ and TeO$_2$ back to CdTe (**Figure 5 a-c**).[40] Previous reports of AA sensing relied on single irradiation mode in which variation in signal due to the fluctuation of incident light, probe concentration, as well as environmental effects were a common problem.[79,80] The current method due to the ratiometric mode is unaffected by these limitations.

Diestra and colleagues in 2017 reported L-cysteine capped ZnS:Mn QDs for detection of DA at room temperature. The QDs display a prominent orange emission band peaking at ~598 nm, which is strongly quenched upon addition of DA in alkaline medium. These results are explained in terms of a pH-dependent electron transfer process, in which the oxidized DA quinone functions as an efficient electron acceptor (**Figure 5 d,e**). The sensor exhibits a linear working range of ~0.15 to ~3.00 µM and a detection limit of ~7.80 nM.[41] This method can be used for sensing DA in urine and other body fluids that mainly consist of amino acids, glucose, and ions. Compared to previous reports for detecting DA by colorimetry,[81] chemiluminescence,[82] gas[83] and liquid chromatography,[84] and electrochemical[85] based approaches, the current work offers short operation time, less overlapping signals, easy method of extraction, and low interference from other biological molecules.

In addition to various sensing applications of metal based QDs for bioanalytes in solution, several reports on their *in vitro*[86,87] and *in vivo*[88] applications have been published. For example, Medintz *et al.* reported the use of QD/DA bioconjugates for intracellular pH sensing in COS-1 cells.[89] The sensor consisted of CdSe/ZnS QDs ($\lambda_{em}$: 550 nm) coated with PEG-modified dihydrolipoic acid (DHLA) ligands to make it hydrophilic. These hydrophilic DHLA-PEG QDs were further linked to DA labelled peptides (**Figure 5f**). At low pH, hydroquinone acts as a poor electron acceptor resulting in low QD PL quenching.

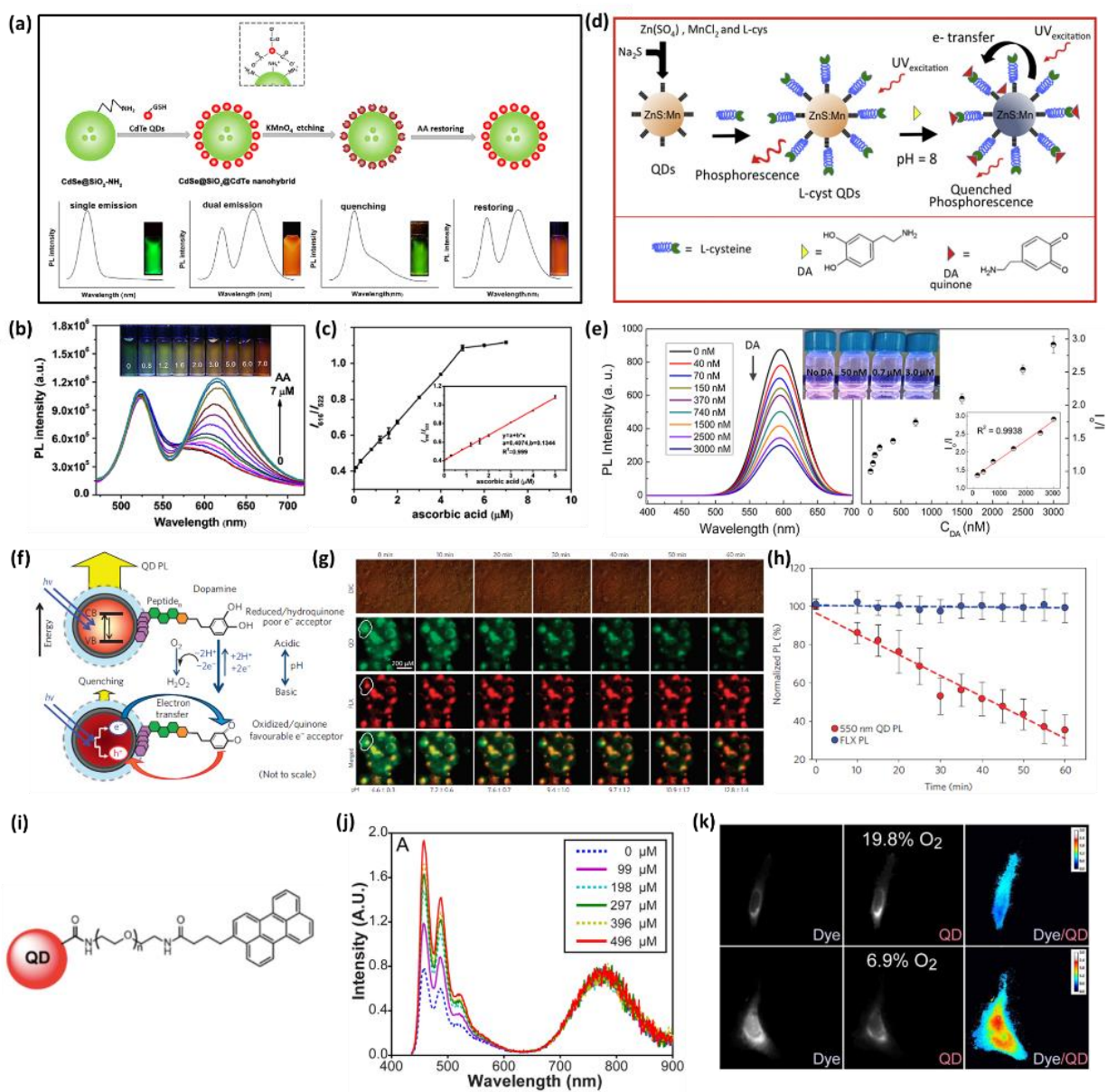

**Figure 5.** (a) Nanohybrid preparation and the mechanism of AA detection. (b) PL spectra of the KMnO$_4$ (7.0 μM) oxidized CdSe@SiO$_2$@CdTe NPs (50 μg mL$^{-1}$) at different concentrations of AA. (c) PL intensity ratio (I$_{616}$/I$_{522}$) of the KMnO$_4$ oxidized nanohybrid vs. AA concentration.[40] (d) Sensor design and selective Dopamine (DA) detection through phosphorescence quenching. (e) Phosphorescence emission spectra and Stern-Volmer plots of l-cys capped ZnS:Mn QDs at different concentrations of DA.[41] (f) Schematic showing mechanism of pH sensing with QD/DA bioconjugates. (g) DIC and fluorescence micrographs of COS-1 cells co-injected with 550 nm-emitting QD–DA conjugates and FLX nanospheres. (h) QD and FLX PL intensities collected from the images shown in (g) vs. time.[89] (i) Scheme of a glucose sensitive AgInS$_2$/ZnS QD linked to oxygen-sensitive perylene dye derivative. (j) Emission spectra of sensor with increasing glucose concentration. (k) Ratiometric response of the sensor to hypoxia in HeLa cells. [94]

As pH increases, ambient $O_2$ in the buffer oxidizes DA, causing an increase in quinone concentration that acts as a favorable electron acceptor in close proximity to the QD. This produces higher quenching efficiencies in a magnitude directly proportional to the amount of quinone. The authors took advantage of the fact that DA increases the quenching rate of the QDs as it can act as an electron acceptor and can cause FRET.[90] They showed that at low pH the catechol groups (in hydroxyquinone form) of the DA are unable to quench the QDs, but with an increase in pH, the hydroxyquinone form is converted to quinone due to oxidation. The modified DA in quinone form was shown to quench the fluorescence in a pH dependent manner. The authors also discussed intracellular pH sensing capability of the QD system, where they used a fluorophorex 20 nm nanosphere (FLX) as reference fluorophore, with fluorescence emission at 680 nm. Both types of particles were inserted in COS-1 cells using microinjection followed by incubating the cells in nystatin containing solution of PBS at pH 11.5. Nystatin was used to form micropores that allowed the exchange of $H^+$/OH with the extracellular environment. The cells were then imaged using fluorescence microscopy at different time points to observe the change in fluorescence emission (**Figure 5g**). As expected, the fluorescence from the DA-QDs decreased over time due to an increase in intracellular pH, while emission due to FLX remained unchanged (**Figure 5h**). The possibility of reversibility of the sensor was not investigated, which is a very important aspect. Also, in most biological events a decrease in pH is more common rather than an increase.

In the past oxygen-sensitive QDs were explored to measure dissolved oxygen concentration *in vitro*[91,92] and *in vivo*.[93] Most of the reported semiconducting QDs based systems don't show intrinsic sensitivity towards oxygen and an additional oxygen sensitive fluorophore has to be used, which should have a long fluorescence lifetime. Shamirian *et al.* reported detection of hypoxia using $AgInS_2$/ZnS QD linked to perylene dye[94] (**Figure 5i,j**). The sensors were microinjected into HeLa cells allowing sensing of hypoxia intracellularly by means of fluorescence microscopy (**Figure 5k**). This sensing platform allowed ratiometric sensing where the $AgInS_2$/ZnS QD acted as a reference that can be excited at 365 nm to get emission at 775 nm. The $O_2$ sensitive fluorescence came from an oxygen-sensitive perylene dye derivative which showed excitation and emission at 427 nm and 472 nm, respectively. The use of less cytotoxic $AgInS_2$/ZnS QDs is the main advantage of this system as conventional CdSe based QDs are intrinsically toxic. Although the authors claimed that their system is better suited for *in vivo* applications due to the use of IR emission from the QDs, the $O_2$ sensitive emission in the visible range would still be vulnerable to scattering from the tissue.

## 2.2 Metal nanoclusters

Studies on fluorescent metal nanoclusters (NCs) and fluorophore linked metal NPs in sensing applications have been explored for the past few years. Applications of fluorophore linked metal NPs species, like gold and silver NPs, have been reported in large numbers,[95–98] where sensitivity was achieved by using them either as a quencher to an organic fluorophore or as a plasmon based sensor. Due to their larger size, they do not show fluorescence emission, but their application in combination with fluorescent QDs and organic fluorophores has been explored with interest for the past couple of years.[96,99–101] Metal nanoclusters, on the other hand, are intrinsically fluorescent and are currently explored as a relatively new class of fluorophores compared to organic small molecule dyes, semiconductor QDs or fluorescent proteins molecules. They offer superiority over organic fluorophores as they are not vulnerable to photobleaching, which is a big limitation in the case of organic fluorophores which limits their range of applications.[102,103] As another alternate, semiconductor QDs showcase tunable fluorescence with a photo-stable emission, however, their large physical size and intrinsic cytotoxicity is a challenging problem, which limits their use for *in vivo* applications.[104] Therefore, metal NCs with strong photoluminescence, appreciable photostability, sub-nanometer size, low cytotoxicity, and tunable fluorescence offers an ideal platform for developing biological sensors.[105,106]

Generally, metal NCs are synthesized by reduction of metal ions in the presence of suitable reducing agents, similar to the synthesis of metal NPs. However, under such conditions, metal NCs are prone to strongly interact with each other causing irreversible aggregation which reduces their surface energy, therefore, resulting in the formation of large NPs. Thus, a suitable stabilizing scaffold is a necessity to produce metal NCs. The nature of the scaffold is responsible not only for their sizes but also for their fluorescence properties. Different kinds of scaffolds which can be used for the synthesis of metal NCs include DNA oligonucleotides, peptides, proteins, dendrimers, and polymers.

Use of metal NCs for labelling of biomolecules, such as oligonucleotides, peptides, and proteins has resulted in the novel application for specific biosensing and bioimaging. Such bioconjugations are commonly based on passive adsorption, multivalent chelation, and covalent-bond formation. As an example, Nandi *et al.* reported multifunctional gold nanoclusters (AuNCs) using self-assembled bovine serum albumin (BSA) nanofiber as a scaffold.[47] The AuNCs stabilized by the amyloid fibril showed appreciable enhancement in fluorescence emission and a redshift of 25 nm compared to its monomeric protein counterpart (BSA-AuNC). The authors demonstrated that these AuNCs could be utilized for cysteine sensing both *in vitro* and *in vivo* even in presence of homocysteine (Hcy) or glutathione (GSH), which often interfere in its detection as observed when using other techniques like high performance liquid chromatography, post column derivatization and a spectrophotometric assay using Ellman's reagent.[107,108]

In another instance, Govindaraju and co-workers reported fluorescent AuNCs for DA sensing in cerebrospinal fluid.[109] They also used BSA for stabilizing the AuNCs, where the fluorescence emission intensity of BSA-AuNCs was quenched upon addition of varied concentration of DA via electron transfer mechanism. The detection range of DA was 0 to 10 nM with a detection limit of 0.622 nM and 0.830 nM in phosphate buffer saline and cerebrospinal fluid, respectively. The sensor could be potentially used in studies on diseases like Parkinson's and Alzheimer disease. This work offered the lowest limit for detection of DA compared to previously reported fluorescence based methods.[110–112]

Using lysozyme type VI as stabilizing scaffold, Chen *et al.* reported GSH sensitive $Au_8$ cluster (consisting of 8 gold atoms) involving reduction and interaction of $Au^{+3}$ by proteins to generate $Au_8$ NCs (**Figure 6a,b**). The fluorescence emission of the $Au_8$ cluster was centered around 455 nm when excited at 380 nm.[48] Presence of GSH can etch the $Au_8$ NCs and thus quench their fluorescence. The cluster showed a high QY of 56% and two fluorescence lifetimes. Because GSH can induce the core-etching of $Au_8$ clusters, (Lys VI)-stabilized $Au_8$ clusters were applied to detect GSH in a single drop of blood with a detection limit of 20 nM (signal to noise ratio of 3). This work for the first time demonstrated the protein-directed synthesis of blue emitting AuNCs with high QY which can be used for sensing of biomolecules. The QY of $Au_8$ clusters was significantly higher than previously reported protein-stabilized AuNCs[49,113–115] (QY<10%), as proteins can screen the AuNCs against the polar solvent, leading to an increase in fluorescence.

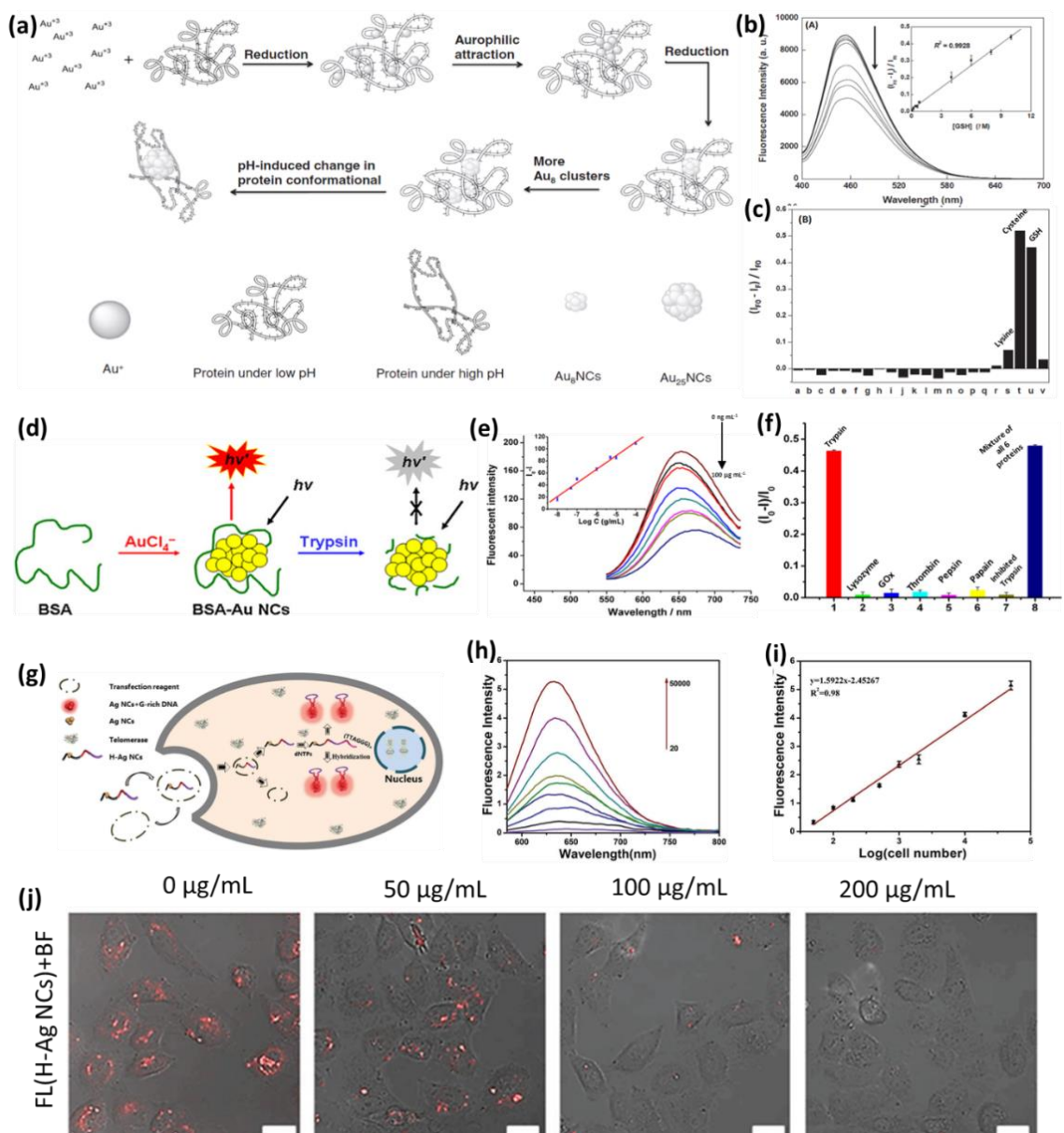

**Figure 6**. (a) Growth mechanism for the synthesis of $Au_8$ and $Au_{25}$ clusters for GSH sensing. (b) Fluorescence spectra of (Lys VI)-stabilized $Au_8$ clusters with increasing concentrations of GSH. (c) Relative fluorescence intensities at 455 nm of (Lys VI) solution-stabilized $Au_8$ clusters in presence of different amino acids.[48] (d) Schematic of sensing trypsin using BSA-AuNCs. (e) Emission spectra of BSA-Au NCs (5 μM) with different trypsin concentrations Inset: Fluorescence intensity decrease ($I_0 - I$) vs. concentration of trypsin (Log scale). (f) Comparison of the relative fluorescence intensity decrease of BSA-Au NCs brought by different enzymes (10 μg mL$^{-1}$).[116] (g) Schematic of H-Ag NC probe for intracellular telomerase activity imaging. (h) Fluorescence intensity for different HeLa cell numbers. (i) Fluorescence intensity vs. HeLa cell numbers (Log scale). (j)

Confocal images of HeLa cells transfected with H-Ag NCs at different concentrations of EGCG (Scale bars: 25 µm).[117]

Although the AuNCs were sensitive for GSH they were also found to be sensitive towards cysteine, thus the interference from cysteine in the biological system can perturb the applicability of the developed system for selective sensing of GSH (**Figure 6c**).

Protease sensing by means of AuNCs was reported for the first time by Hu *et al.* in 2012.[116] The authors synthesized BSA stabilized AuNCs for detection of trypsin. The sensing range was 0.01-100 µg mL$^{-1}$ with a detection limit of 2 ng mL$^{-1}$. The working principle of the BSA-AuNCs was based on the cleavage of BSA scaffold in presence of trypsin causing the deprotection of AuNCs, which ultimately results in a decrease of fluorescence emission due to quenching by the surrounding solvent (**Figure 6d,e**). The cross sensitivity of the platform was assessed against other proteolytic enzymes like lysozyme, glucose oxidase, thrombin, papain, and pepsin, but no obvious decrease in fluorescence was observed **(Figure 6f)**. As a potential biomedical tool, the BSA-AuNCs were used to test trypsin in urine samples and found to be working as expected.[118] Notably, before this report, the applications of protein templated fluorescent AuNCs in sensing applications were mainly based on the fluorescence quenching effect of AuNCs due to ionic interactions ($Hg^{2+}$, $Cu^{2+}$, $CN^-$)[45,114,119,120] or through the oxidation of Au-S bond between Au and protein templates by hydrogen peroxide ($H_2O_2$).[121]

In 2018, Huan *et al.* reported a very sensitive silver NC based fluorescent sensor for quantifying intracellular activity of telomerase.[117] Telomerase sensing is important as its activity can be used as a diagnostic and prognostic biomarker of cancer. In addition, telomerase is also considered as a potential therapeutic target for treatment of cancer. To this end, the authors developed AgNCs containing three functional DNA scaffolds (H sequence): a poly (cytosine) and a T5 loop, a nine base complementary sequence that could hybridize with the telomerase elongation product, and the telomerase substrate primer (**Figure 6g-i**). The mechanism of sensing involved inserting the probes into the cell using a transfection agent, where the telomerase substrate present on the AgNCs can be extended to generate several repetitive TTAGGG sequences. The extended sequence in turn can hybridize with the nine base complimentary sequence TAACCCTAA by forming a self-hairpin structure. The presence of G rich sequence in vicinity of the AgNC enhanced its fluorescence emission, which was used as an indicator of intracellular telomerase activity. The variation in telomerase activity in response to different concentration of telomerase inhibiting drugs (EGCG) was also evaluated, where a correlation between increasing EGCG concentration and decrease in intracellular fluorescence due H-Ag NCs was observed **(Figure 6j)**. The most interesting feature of this sensing method is its real time application and sensitivity, which allows its use in clinical diagnosis and testing of newly discovered telomerase inhibiting drugs.

In another report, sensing of $H_2O_2$ by means of horseradish peroxidase (HRP) functionalized fluorescent AuNCs was demonstrated by Wen *et al.*[49] The fluorescence of the HRP-AuNCs was quenched quantitatively by addition of $H_2O_2$. The quenching effect was linear over the range of 100 nM – 100 µM with a detection limit of 30 nM and signal to noise ratio of 3. The authors also explored the possible reason for $H_2O_2$ sensitivity in the HRP functionalized AuNCs. Observation of HRP-AuNCs by TEM showed the presence of HRP scaffold in the absence of $H_2O_2$. On the other hand, in presence of $H_2O_2$, the size of the AuNCs increased indicating removal of HRP scaffold and aggregation of AuNCs causing a reduction in fluorescence emission. However, for optimum working, HRP-AuNCs required alkaline pH and a temperature of 25°C. Thus, biological applications involving relatively low pH values and high temperature may limit its functioning.

An example of silver NCs based sensors include the work by Lan *et al.* where the authors prepared a fluorescent functional oligonucleotide stabilized silver NC and employed it as a probe for detecting single nucleotide polymorphism (SNPs) in the gene coding for fumarylacetoacetate hydrolase (FAH).[122] The functional oligonucleotide-stabilized silver NCs (FFDNA-AgNCs) were synthesized through $NaNH_4$ mediated reduction of $AgNO_3$ in the presence of fluorescent base motif $C_{12}$ containing the DNA sequence 5'-$C_{12}$-CCAGATACTCACCGG-3', which can recognize the FAH gene. The fluorescence from the FFDNA-Ag NC probe was observed to be sensitive towards the presence of perfect match DNA ($DNA_{pmt}$) in the solution, which causes an increase in fluorescence emission intensity. Under optimal conditions (150 mM NaCl, 20 mM phosphate solution, pH 7.0), the fluorescence ratios of the FFDNA-Ag NC probes in the presence and absence of $DNA_{pmt}$, plotted against the concentration of $DNA_{pmt}$, was linear over the range 25–1000 nM ($R^2$ = 0.98), with a detection limit of 14 nM (Signal-to-noise ratio of 3). This method of employing metal nanoclusters for DNA detection is unique as it is very challenging to replace a concentrated templated DNA around the cluster by a functional DNA or other recognition molecules. Here the authors circumvented this problem by using an additional segment of DNA linked to the target complimentary sequence. This additional segment ($C_{12}$) was used as a scaffold for stabilizing the AgNCs, as it was observed in previous reports that $Ag^+$ ions are known to interact selectively with the DNA heterocyclic bases of cytosine ($C_{12}$ motif), which induces nonplanar and tilted orientations of DNA.[123–125]

## 2.3 Silicon based nanosensors

In general, silicon based NPs such as silica NPs (SiNPs) are intrinsically nonfluorescent due to the indirect band gap[14], whereas few groups have reported intrinsically fluorescent silicon NPs.[126–128] In most of the reports the fluorescence is imparted either by linking a fluorescent dye to the particle surface[129] or by embedding it inside the particle. SiNPs are commonly synthesized by a method devised by Stöber *et al.* in 1968, where tetraethyl orthosilicate (TEOS) is first hydrolyzed followed by polymerization in presence of ammonia in an alcoholic solution, which acts as a catalyst.[130] A wide variety of fluorophores can be incorporated into the SiNPs using this method. Alternatively, the method of reverse emulsification can be used to produce SiNPs smaller than 100 nm.[131–133] In this method, a mixture of oil-water and the surfactant is used to create nanometer sized stable water droplets that can act as nanoreactors for silica particle nucleation and growth.

In the year 2017, Kim and colleagues reported $H_2O_2$ sensing using SiNPs. The dye peroxy lucifer 1 (PL1), was immobilized onto SiNPs (PL1-$SiO_2$) for the detection of intracellular $H_2O_2$.[53] The mechanism of sensing is based on the modulation of internal charge transfer properties of the probe with a change in concentration of $H_2O_2$. It triggers chemo selective cleavage of the boronate-based carbamate protecting group, resulting in internal charge transfer (ICT)-induced redshifts in emission maxima. In ratiometric analysis (F540/F470), a linear correlation between the $H_2O_2$ concentration and carbamate-derived emission intensities was observed using PL1-$SiO_2$. One of the limitations of the developed system could be its irreversible sensing where it can only sense an increase in $H_2O_2$ concentration as the change in fluorescence involves bond breaking.

Yu *et al.* demonstrated a mesoporous SiNPs (MSNs) based sensor for selective detection of Hcy from biothiols and other common amino acids.[52] In this work, the anthracene nitroolefin compound was placed inside the MSNs and used as a probe for thiols (**Figure 7a-c**).

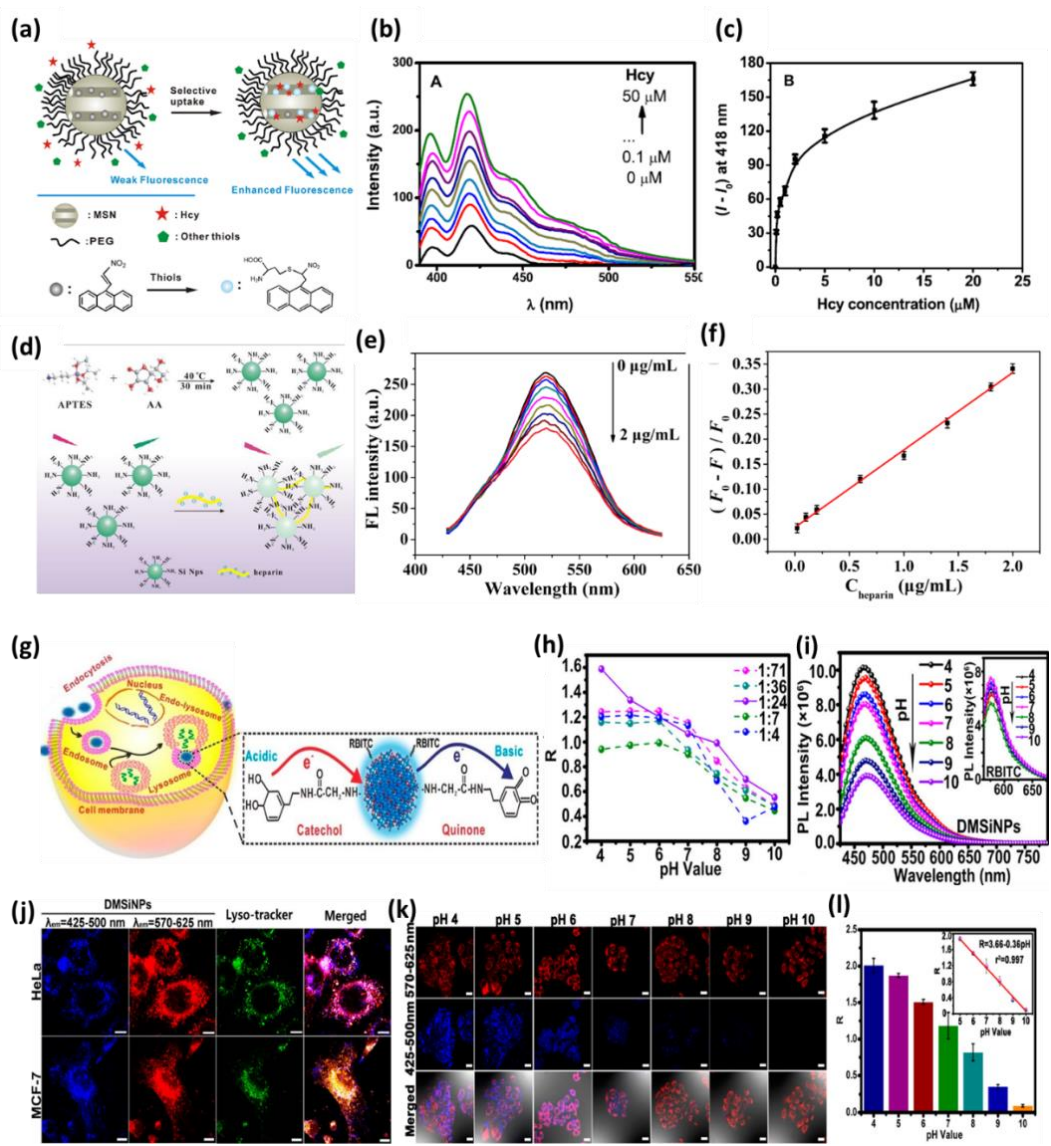

**Figure 7**. (a) Schematic of the MSN-based sensor (probe–MSN–PEG) and its selective detection of Hcy. (b) Fluorescence spectra of the probe–MSN–PEG (0:5 mg mL$^{-1}$) at different concentrations of Hcy. (c) Increment of fluorescence intensity (at 418 nm) of the probe–MSN–PEG (0.5 mg mL$^{-1}$) as a function of Hcy concentration. [52] (d) One-step synthetic strategy of the silicon NPs for heparin detection, (e) Fluorescence spectra of the silicon NPs at various concentrations of heparin. (f) Fluorescence quenching efficiency ($F_0$ - F)/$F_0$ vs. concentration of heparin.[134] (g) Schematic for cellular internalization of DMSiNPs. Inset: Charge-transfer mechanism of DMSiNPs in acidic or basic conditions. (h) Fluorescence intensity ratio (R = $I_{465}/I_{575}$) vs. pH values for DMSiNPs (0.48 mg mL$^{-1}$) at different molar feed ratios of RBITC/DA (1:4 to 1:71). (i) PL spectra of DMSiNPs (0.48 mg mL$^{-1}$) at different molar feed ratios of RBITC/DA (1:24); $\lambda_{exc}$= 405 nm. Inset: PL spectra under same

conditions with $\lambda_{exc}$ = 543 nm. (j) Confocal images of intracellular DMSiNPs in live HeLa and MCF-7 cells co-stained with lysotracker. (Scale bars-HeLa cells:7.5 μm; MCF-7 cells: 10 μm). (k) DMSiNPs-based Intracellular pH calibration (Scale bars: 25 μm). (l) Fluorescence intensity ratio (R = $I_{465}/I_{575}$) of SiNPs  vs. lysosomal pH values. [54]

In addition, hydrophilic polyethylene glycol (PEG 5000) molecules were covalently bound to the MSN surface and used as a selective barrier for Hcy detection via different interactions between biothiols and the PEG polymer chains. Due to the existence of PEG chains on the particle surface, the amount of Hcy that diffuses into the mesopores is much more than that of other thiols. The authors showed that this sensor could discriminate Hcy from the two structurally similar low-molecular-mass biothiols (glutathione and cysteine) and other 19 common amino acids.

In another work, Ma *et al.* reported sensing of heparin using water-soluble silicon NPs that were synthesized by a one-step method using 3-aminopropyl triethoxysilane (APTES) as a silicon source and L-ascorbic acid (AA) as the reducing reagent.[134] APTES is reduced by AA to generate amino functionalized fluorescent silicon NPs. The fluorescence of the silicon NPs was sensitive to the presence of heparin, as it caused aggregation due to hydrogen bonding leading to quenching (**Figure 7d-f**). The linear range of heparin sensing was 0.02 to 2.0 μg $mL^{-1}$, with a detection limit of 18 ng $mL^{-1}$ (equal to 0.004 U $mL^{-1}$). In comparison to various other previously reported nanoparticle based probes,[135–137] it has a significantly lower limit of detection, making it more promising for clinical use.

Chu *et al.*, synthesized dual modified pH sensitive silicon NPs (DMSiNPs) by using pH-sensitive DA and pH-insensitive rhodamine B isothiocyanate (RBITC).[54] The synthesized NPs were then utilized for assessing the lysosomal pH change mediated by nigericin in HeLa and MCF 7 cell lines (**Figure 7g-l**). Notably, DMSiNPs showed a very high resistance towards photo quenching compared to the lysotracker (DND 26). Using the sensor, the authors were able to observe two consecutive steps of cellular growth, i.e., alkalization lag phase and logarithmic growth phase, by recording the pH change. These sensors have a potential of providing important information about the dynamic process of pH fluctuations. The excellent photostability and low cytotoxicity makes this class of pH nanosensors a very suitable option for investigating intracellular behavior.

## 3.  ORGANIC NANOSENSORS
### 3.1 Carbon-based nanosensors
For the past few decades, carbon based nanosensors have emerged as a competitive substitute for semiconducting QDs. Very small size (less than 10 nm) and fluorescence are innate properties of CDs.[138] Their structure mainly consists of carbons with $sp_2/sp_3$ hybridization, in addition to oxygen/nitrogen-based groups or polymeric aggregations. Their simple composition makes them less toxic and provides special photophysical properties which are tailorable. They can be broadly classified into three categories: graphene QDS (GQDs),[139] carbon QDs (CQDs),[140,141] carbon dots (CDs).[138] The origin of fluorescence in carbon based nanosensors can differ due to variation in precursors and reaction conditions while semiconducting QDs, rely totally on quantum confinement. Different groups have associated this property of CDs to different phenomena like quantum confinement,[23] PL centers,[25] and presence of fluorescent organic molecules on the surface on the surface of CDs.[142]

GQDs contain graphene (one or more layers) with chemical groups like hydroxyl, carboxyl, primary amines etc. attached to their edges.[139] Their lateral dimensions are larger than their height making them anisotropic. On the contrary, CQDs are always spherical. Various reports mention carbon NPs without a crystal lattice while others have reported CQDs with an apparent crystal lattice structure. Because of this variation in properties, the PL center can vary within CDs of different kinds, offering diverse approaches to fabricate them. The common approaches for the synthesis of CDs are classified into "top-down" and "bottom-up".[143] The top-down synthesis involves chemical or physical cutting processes of macroscopic carbon structures, such as carbon nanotubes (CNTs), graphene, suspended carbon powders, etc. Techniques such as electrochemistry, chemical oxidation, and laser irradiation are the commonly reported methods for cutting the large carbon structures. In the "bottom-up" approach, synthesis of CDs from smaller organic molecules or polymers, and modification of surface functionality or passivation of CDs, is achieved using organic precursors under certain conditions. In this case, microwave,[144] heat,[145] and ultrasonic wave[146] are among the primary methods used for molecular structure transition leading to the formation of CDs. Numerous publications have reported the application of carbon nanosensors for sensing[147–149] with number of publications still increasing. Most of developed systems focused on sensing of pH,[55,60,150] metal ions,[151] glucose,[58] and various enzymes.[57,152] In addition, there are plenty of other reports on the detection of pesticides[153–156] and important analytes of interest such as common heavy metal pollutants.[157–159]

As an example of glucose sensing using CDs, Shen *et al.* reported a one-step fabrication of boronic acid functionalized CDs. A one-step hydrothermal reaction was used with phenylboronic acid as the sole precursor of the reaction. In presence of glucose, the CDs assembled together causing a decrease in fluorescence emission by quenching. The added glucose molecules selectively lead to the assembly of CDs due to covalent binding between the cis-diols of glucose and boronic acid on the CD surface. It was possible to quantify glucose in the range of 9–900 μM.[160,161] The authors also performed real sample assays which required minute quantities (20 μL) of serum samples. This CD based glucose sensor was found to be almost 10–250 times more sensitive than other previously reported boronic acid based fluorescent nanosensing systems. Thus, it is one of the most sensitive boronic acid-based glucose sensors that can even differentiate between both analogs of glucose in addition to other interfering substances in the blood.

In 2016, Li *et al.* reported the synthesis of nitrogen-doped carbon dots (N-CDs) for sensing of alkaline phosphatase (ALP),[57] which is an important biomarker associated with various diseases like liver dysfunction, diabetes, breast and prostate cancer. They used ethanediamine and catechol as precursors for one-pot synthesis of N-CDs. The mechanism of sensing ALP involved the use of p-nitrophenylphosphate (PNPP) as a substrate for the ALP, which generated p-nitrophenol that ultimately decreased the excitation of CDs due to inner filter effect (IFE). Because of the competitive absorption, the excitation of CDs was significantly weakened, resulting in the quenching of CDs (**Figure 8a-c**). This IFE-based sensing strategy showed a good linear relationship from 0.01 to 25 U L$^{-1}$ (R$^2$ = 0.996) and provided an exciting detection limit of 0.001 U L$^{-1}$ (Signal-to-Noise ratio: 3). Further studies were conducted to analyze the enzyme concentration in serum samples, where as little as 10 µL of serum sample was required for the test. Compared to previous reports on CDs, this work stands out as it showed a high QY of 49% and a higher sensitivity of 0.001 U L$^{-1}$, which is 1-2 orders of magnitude more sensitive. The lack of any covalent modification of the carbon dot is another advantage of this system, which enhances its reproducibility.

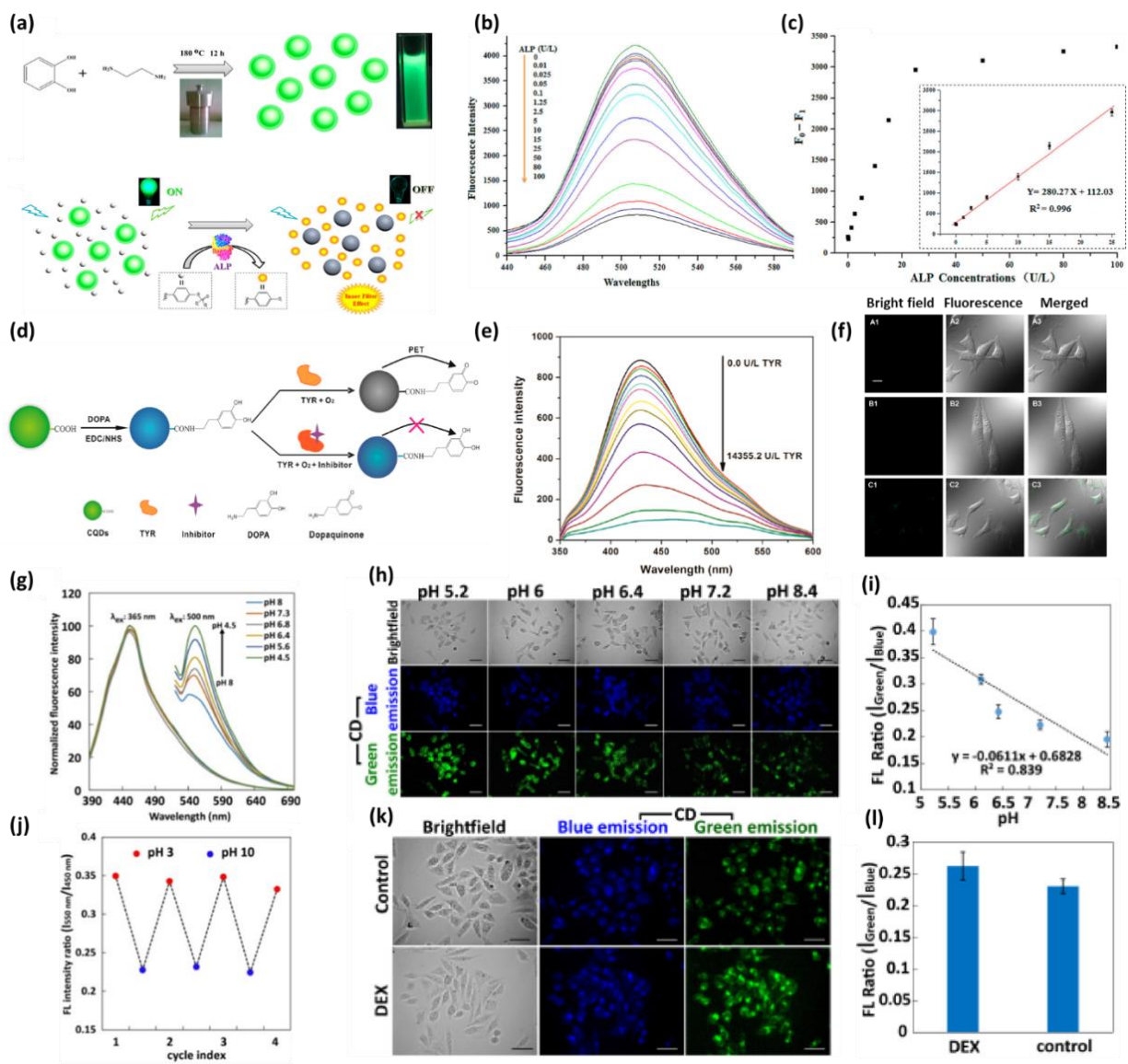

**Figure 8.** (a) One-pot synthesis of N-CDs using ethanediamine and catechol. (b) Fluorescence spectra of N-CDs at various concentrations of ALP with 1 mM PNPP and 0.1 μM MgSO$_4$ (c) F$_0$ − F$_1$ vs. ALP concentration. Inset: Corresponding linear relationship from 0.01 to 25U L$^{-1}$.[57] (d) Schematic for the detection of TYR activity. (e) Fluorescence spectra of Dopa-CQDs as a function of TYR levels. (f) Confocal images of B16 cells under different conditions: (A1–A3) intact cells; (B1–B3) treated cells with 100 μg mL$^{-1}$ Dopa-CQD conjugate; (C1–C3) pre-treated cells with 200 μM Arbutin and then incubated with 100 μg mL$^{-1}$ Dopa-CQD conjugate. Scale bar: 20 μm.[162] (g) Emission spectra of pH sensitive CDs at λ$_{exc}$ 365 nm and 500 nm. (h) Cyclic variation in fluorescence emission intensity ratio (I$_{550}$/I$_{450}$) of CDs in PBS at pH values 3.0 and 10.0 (i) Fluorescence microscopy images of HeLa cells fixed at different pH values. (j) Plot of fluorescence intensity ratio (I$_{green}$/I$_{blue}$) vs. pH (k) Fluorescence microscopy images of HeLa cells with and without dexamethasone (DEX) treatment. (I) Fluorescence ratio change corresponding to intracellular pH changes in DEX treated and untreated cells (Scale bars: 50 mm).[60]

The CDs have also been explored in various other cell-based sensing studies.[163,164] Chai *et al.*, reported DA functionalized CQDs (Dopa-CQDs) for assessing tyrosinase (TYR) activity and screening of its inhibitors based on photoinduced electron transfer (PET) between CQDs and dopaquinone moiety.[162] The DA functionalized on the surface of the CQDs were prone to oxidation to generate dopaquinone, where TYR acts as a specific catalyst (**Figure 8d-f**). The dopaquinone, in turn, causes quenching of CQD by PET. This process was utilized to correlate TYR activity with its fluorescence quenching efficiency, where a more active TYR will cause more fluorescence quenching in the Dopa-CQD complex. The assay covered a broad linear range from 7.0 U/L to 800 U/L. Studies on TYR inhibitors was also performed, where arbutin, a typical inhibitor of TYR, was chosen as an example to assess its function of inhibitor screening. In presence of arbutin, the fluorescence quenching was reduced. The biocompatibility and sensitivity of the Dopa-CQD system were also demonstrated by measuring intracellular tyrosinase level in melanoma cells. A major limitation of Dopa-CQDs reported by other groups as well, is the quenching of its fluorescence at pH values more than 9.0. In addition, the emission changes very abruptly at acidic pH values which can interfere with its specific use in TYR sensing.

In another instance, pH sensing using CDs has been recently reported by Chandra *et al.*, where they have synthesized a ratiometric fluorescent CD from *Agaricus bisporus* using a hydrothermal reaction.[60] The CDs were differentially sensitive towards change in pH, where excitation using 365 nm resulted in pH insensitive emission at 450 nm, while excitation at 500 nm gave rise to emission at 550 nm which was sensitive towards pH change. The ratio of these two emissions was used to estimate the intracellular pH in HeLa cells. The sensitivity range of the CDs was pH 4-10. The authors showed the effectiveness of the developed CDs in estimating acidification of intracellular pH in HeLa cells treated with dexamethasone, which is known to cause intracellular acidification. A more acidic pH is observed in cells treated with DEX compared to untreated control cells (**Figure 8g-l**). These CDs offered one step synthesis and minimal processing. Compared to other ratiometric pH probes, these CDs were able to sense pH ratiometrically without using any reference fluorophore, which reduced the additional modification steps, showing robustness of the CDs fluorescence emission for a strong repeated cyclic pH fluctuation.

### 3.2 Polymeric NPs
Polymeric NPs represent another group of nanosensors, which have been used very often for optical diagnostics, such as biomarker analysis, cancer diagnosis, diagnostic imaging, and immunoassays. The dye which is used for sensing is generally prone to limitations like photobleaching and low emission intensity. However, in polymeric NPs the dye is embedded within the polymeric matrix thereby increasing its photostability due to "protective" effect of the polymer.[165,166] A fine strategy of covalently coupling the dye to the polymer is often required to avoid dye leakage, which is a possibility with entrapped dyes. Another advantage of polymeric NPs-based systems is their ability to encompass hundreds of dye molecules which increases the brightness of emission and hence the sensitivity.[21] In some cases, the hydrophobic microenvironment created within polymeric NPs can even improve the brightness of certain fluorescent dye molecules.[167,168] Polymeric NPs also offer the possibility to include a reactive functional group along the polymer backbone allowing labelling with another molecule for a more specific function.

The synthesis of polymeric nanosensors can be achieved either by polymerization of monomers or by processing of preformed polymers.[169] In the latter case, the most commonly employed techniques are solvent evaporation,[170] salting-out,[171] nanoprecipitation,[172] dialysis[173] and supercritical fluid technology.[174] Formation of polymeric NPs from the monomers has been achieved by a variety of methods, such as dispersion,[175] precipitation,[176] and interfacial polymerizations,[177] etc. For diagnostic applications, the most frequently used techniques are emulsion and living free radical polymerization.[178]

A noteworthy report on polymeric nanosensors includes work of Wang and co-workers, where they showed organic fluorescent nanocomposite for DNA detection.[179] The sensor consisted of π-conjugated fluorescent oligomers 4,7-(9,9_-bis(6-adenine hexyl)- fluorenyl)-2,1,3-benzothiadiazole (OFBT-A), which was reprecipitated with 5'-Texas Red-CGGAG-3' (TR-P5) (**Figure 9a-c**). FRET between the OFBT-A and TR-P5 was observed to be differentially sensitive against DNA of different lengths. By using these sensors, the authors were able to monitor the DNA elongation and the cleavage processes, which are essential steps in DNA replication and repair. One of the advantages of this system is its sensitivity only for length of DNA and not on sequence, therefore it avoids sequence dependent interference.

Although detection of DNA length can be used in certain applications, sequence specific detection of DNA is another important aspect which is very essential in diagnostic assessments for genetic mutations for various metabolic disorders. To this end, various reports in the past have mentioned about fluorescent probes that can sense extremely low concentrations (about few pM) of nucleic acids (NAs) with specific sequence. But they have to rely on molecular multiplication using enzymes.[180–182] In an effort to achieve one step amplification without using an enzyme, Wang et al. reported cationic conjugated polymers that can sense sequence-specific hybridization with improved detection limits down to the 10 pM range.[183] However, it suffered from non-specific interaction with other NA duplexes and proteins, and strong dependence on the ionic strength. As a better one-step non enzymatic amplification platform for DNA sensing, Melnychuk et al. reported a FRET-based polymeric NPs.[63] The sensor consisted of dye-loaded

poly(methylmethacrylate-co-methacrylic acid) (PMMA-MA) NPs of 40 nm size with conjugated oligonucleotides, where together they acted as nanoantenna (**Figure 9d-f**). The nanoantenna served as a super efficient energy donor in a FRET-based NA detection, where only a few DNA hybridization events at the surface of the NP were able to switch on/off energy transfer from thousands of dye molecules inside the NP. The nanoprobe operated in solution and on surfaces with NA detection limits of 5 and 0.25 pM, respectively, which is 1000–10000-fold lower than that achieved using molecular probes. The universality of this FRET-based detection concept based on functionalizable organic nanoantennas opened a possibility for the development of a broad range of ultrabright probes with amplified detection capabilities.

In another report, Wang *et al.* showed a turn-on fluorescent probe for sensing cysteine,[62] which is a biothiol and plays crucial roles in maintaining the redox homeostasis and active endogenous antioxidants. They synthesized fluorescent organic NPs (FONs) using NIR emissive luminogen (A1), where A1 consisted of tetraphenylethene, diketopyrrolopyrrole, and nitroolefin moieties. The A1 molecule was encapsulated in pluronic F-127 polymer using nanoprecipitation method and after hydrophobic interaction with surfactant pluronic F127, A1 dispersed well in water and aggregated into cores to form A1-F127 FONs. In the normal state, the fluorescence of A1 is quenched by nitroolefin moieties. In presence of biothiols, the nitroolefin molecules react with them causing fluorescence recovery of the A1 unit. The diffusion barrier of hydrophilic polyethylene oxide chains on the surface of the particle selectively allowed sensing of cysteine residues compared to other biothiols.

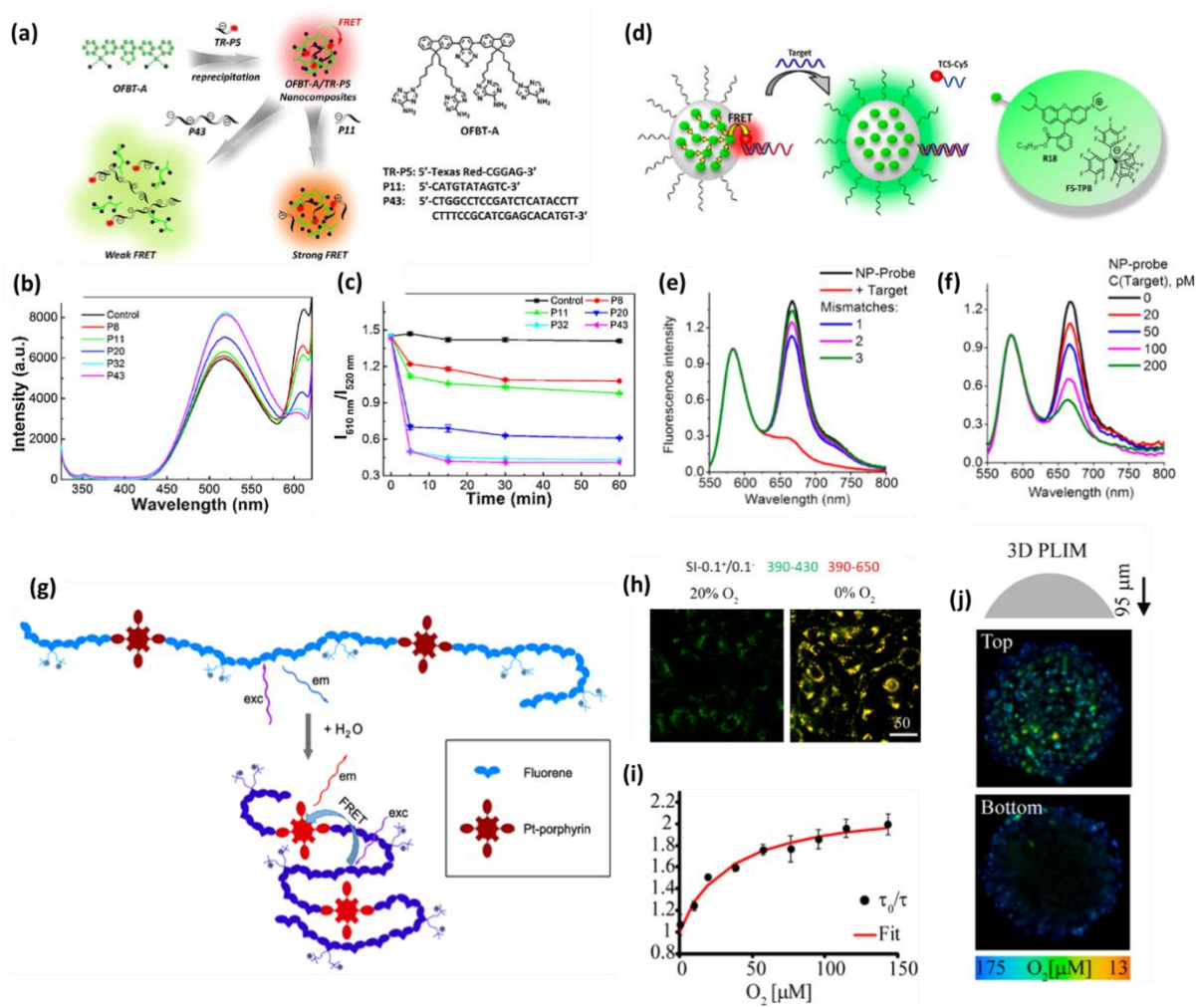

**Figure 9.** (a) Schematic of OFBT-A/TR-P5 nanocomposites for DNA detection. (b) Emission spectra of OFBT-A/TR-P5 nanocomposites with different DNA ($\lambda_{exc}$: 315 nm). (c) FRET ratio $I_{610}/I_{520}$ as a function of action time, in the presence of different DNA.[179] (d) Schematic of DNA sensing using dye-loaded NPs bearing noncoding (gray) and capture (dark red) oligonucleotides. (e) Fluorescence spectra of NP-probe (100 pM) with and without survivin NA target (1 nM). (f) The response of nanoprobe (10 pM) to different concentrations of target nucleotide.[63] (g) Conformation of $O_2$-sensitive conjugated polymers in an organic solvent and in NP form after precipitation with water. (h) Ratiometric intensity imaging of NPs in MEF cells exposed to different $O_2$ percentage. Combined polyfluorene (390-430 nm, green) and Pt (II)-porphyrin (390-650, red) intensity. (i) Stern-Volmer plot for phosphorescence lifetime $O_2$ calibration of NPs (SI-0.1$^+$/0.1$^-$) in MEF cells. (j) 3D reconstruction of $O_2$ distribution by PLIM for the SI-0.1$^+$/0.1 stained HCT116 spheroid.[184]

One of the crucial advantages of this nanosensor is NIR emission that can enable sensing in thick tissues, which are known to scatter visible light and reduce the signal output. In addition, the background fluorescence and damage to the tissue is also known to be less using NIR wavelength.

Thus, for steady observation of thiols in the *in vitro* and *in vivo* systems, it can act as an effective model system.

In another instance, Chen *et al.* reported FRET mediated multifunctional sensors for sensing copper and sulfide ions.[61] The NPs were produced by co-polymerization of styrene, vinylbenzylchloride and fluorescent vinylic crosslinking monomer fluorescein-O,O-bis-propene (FBP) in miniemulsion of oil-in-water stabilized with a cationic surfactant. In addition, 1,4,7,10-tetraazacyclododecane (Cyclen) was grafted on to the surface of NPs. Due to specific FRET energy transfer between FBP and the $Cu^{2+}$-Cyclen complex, the NPs displayed highly sensitive (detection limit: 340 nM), and on-off-type fluorescence change with high selectivity toward $Cu^{2+}$. The fluorescence was recoverable by addition of sulfide ions, thus a fully quenched system of cyclen-capped FPNs–$Cu^{2+}$ complex was used for sulfide ion sensing with a detection limit of 2.1 µM. This NP based dual ion sensor can be reversibly switched for multi times by alternative addition of adequate $Cu^{2+}$ or $S^{2-}$. Its relatively wide pH range (pH 4.0-10.0) stability, and excellent long-term photostability for $Cu^{2+}$ detection (≥40 days) in aqueous media, are some of its interesting features. Thus, this approach of sensing $Cu^{2+}$ may reveal a new pathway for selective detection of multiplex analyst in environmental and biological applications.

A recent example of pH-sensitive polymeric NPs includes work reported by Zhao et al. [64] The synthesized poly(3,4-dihydroxyphenylalanine) fluorescent organic NPs (poly (DOPA)-FONPs) showed two emission peaks under single excitation, where both the emissions were sensitive towards pH change. Therefore, the sensor was ratiometrically sensitive towards pH change in the range 4.0-8.0. Due to its low cytotoxicity, the poly (DOPA)-FONPs were also utilized for cell imaging. The special feature of these NPs is their intrinsic fluorescence without the use of any additional dyes, that often requires complex linking reactions. Additionally, excellent biocompatibility of these NPs makes them suitable for long term studies.

An example of real time *in vitro* and *in vivo* sensing using polymeric NPs involves work reported by Dimitriev *et al.*, where they reported conjugated polymer NPs for high-resolution $O_2$ imaging in cells and 3D tissue models.[184] The sensor consisted of a substituted conjugated polymer (polyfluorene or poly(fluorene-alt-benzothiadiazole)) acting as a FRET antenna and a fluorescent reference with covalently bound phosphorescent metalloporphyrin (PtTFPP, PtTPTBPF), which are among popular $O_2$ indicators (**Figure 9g-j**). They produced different types of conjugated polymers (denoted as SI and SII) with a variety of charged groups (+, -, or ±) to form cationic, anionic or zwitter-ionic structures. They found that NPs with zwitterionic charge (SI-$0.1^+/0.1^-$) were more efficient in staining mouse embryonic fibroblast cells (MEF cells) compared to positively charged NPs. Using SI-$0.1^+/0.1^-$ NPs they were able to stain MEF cells exposed to different $O_2$ at 37 °C and treated with an inhibitor of respiration (antimycin A, 10 µM), which showed a clear difference in emission intensity according to $O_2$ concentration. It was also found that SI-$0.1^+/0.1^-$ NPs can efficiently penetrate across multiple cell layers in 3D models of HCT116 and PC12 cells. The flexibility of the platform to tune the charge and spectral properties presents a unique and promising alternate for dissolved oxygen sensing for tissue-based studies *in vitro* and *in vivo*. In addition, the low cytotoxicity of these nanoprobes offers long term sensing capabilities with least interference to the system being studied.

## 4. CONCLUSIONS

NPs-based fluorescent sensors have been the subject of intense studies due to their interesting properties. They have evolved swiftly from just being a brighter alternative to organic fluorophores towards smart sensing probes for the development of better sensing and molecular diagnostic strategies. Over the past few decades, NPs-based fluorescent sensors have been utilized as a valuable tool to monitor sensitive biological reactions and sensing biomolecules in ultra-low concentrations, which was unachievable using traditional organic fluorophores. Despite their proven usefulness, it is important to mention that organic fluorophores can never be undermined. NPs have their own limitations such as relatively larger size and cytotoxicity, which restricts their applications. Therefore, both organic fluorophores and NPs-based sensors should be looked upon as complementary tools in biosensing. In future, more improvement is required in the controlled synthesis of NPs in relation to their consistency, size distribution, large-scale purification and long-term stability. At the same time, in order to promote their clinical translation and application, more preclinical and clinical trials are needed to assess the long-term effects of NPs-based sensors in preclinical models and in humans, as they can interact at a variety of levels (e.g., with proteins, oligonucleotides, organelles, cells, and organs). For *in vivo* applications, extensive research should also be directed towards the development of ease of use and portable imaging equipment with higher tissue penetration, and the creation of computational/modeling tools for automated data analyses of fluorescent images to facilitate more accurate and reliable quantitation of the occurring biological processes in the analyzed systems.

## 5. CONFLICTS OF INTEREST

There are no conflicts to declare.

## 6. ACKNOWLEDGMENTS

The authors gratefully acknowledge the ERC Starting Grant INTERCELLMED (project number 759959) and the project "TECNOMED—Tecnopolo di Nanotecnologia e Fotonica per la Medicina di Precisione" [Ministry of University and Scientific Research (MIUR) Decreto Direttoriale n. 3449 del 4/12/2017, CUP B83B17000010001].

## 7. BIBLIOGRAPHY

1. Herschel, S. J. On a case of superficial colour presented by a homogeneous liquid internally colourless. *Philos. Trans. R. Soc. Lond.* **135**, 143–145 (1845).
2. Valeur, B. & Berberan-Santos, M. N. *Molecular Fluorescence: Principles and Applications*. (Wiley-VCH Verlag GmbH & Co. KGaA, 2012). doi:10.1002/9783527650002
3. Wang, X., Chen, X., Xie, Z. & Wang, X. Reversible Optical Sensor Strip for Oxygen. *Angew. Chem. Int. Ed.* **47**, 7450–7453 (2008).
4. Yin, J., Peng, M., Ma, Y., Guo, R. & Lin, W. Rational design of a lipid-droplet-polarity based fluorescent probe for potential cancer diagnosis. *Chem. Commun.* **54**, 12093–12096 (2018).
5. Toh, K. C., Stojković, E. A., Stokkum, I. H. M. van, Moffat, K. & Kennis, J. T. M. Proton-transfer and hydrogen-bond interactions determine fluorescence quantum yield and photochemical efficiency of bacteriophytochrome. *Proc. Natl. Acad. Sci.* **107**, 9170–9175 (2010).
6. Fehérvári, T. D., Okazaki, Y., Sawai, H. & Yagi, T. In Vivo Voltage-Sensitive Dye Study of Lateral Spreading of Cortical Activity in Mouse Primary Visual Cortex Induced by a Current Impulse. *PLOS ONE* **10**, e0133853 (2015).
7. Albertazzi, L. *et al.* Dendrimer-Based Fluorescent Indicators: In Vitro and In Vivo Applications. *PLOS ONE* **6**, e28450 (2011).
8. Jiang, W., Fu, Q., Fan, H. & Wang, W. An NBD fluorophore-based sensitive and selective fluorescent probe for zinc ion. *Chem. Commun.* **0**, 259–261 (2007).
9. López-Duarte, I., Vu, T. T., Izquierdo, M. A., Bull, J. A. & Kuimova, M. K. A molecular rotor for measuring viscosity in plasma membranes of live cells. *Chem. Commun.* **50**, 5282–5284 (2014).
10. Colom, A. *et al.* A fluorescent membrane tension probe. *Nat. Chem.* **10**, 1118 (2018).
11. Arai, S., Lee, S.-C., Zhai, D., Suzuki, M. & Chang, Y. T. A Molecular Fluorescent Probe for Targeted Visualization of Temperature at the Endoplasmic Reticulum. *Sci. Rep.* **4**, 6701 (2014).
12. Ruedas-Rama, M. J., Walters, J. D., Orte, A. & Hall, E. A. H. Fluorescent nanoparticles for intracellular sensing: A review. *Anal. Chim. Acta* **751**, 1–23 (2012).
13. Wolfbeis, O. S. An overview of nanoparticles commonly used in fluorescent bioimaging. *Chem. Soc. Rev.* **44**, 4743–4768 (2015).
14. Ng, S. M., Koneswaran, M. & Narayanaswamy, R. A review on fluorescent inorganic nanoparticles for optical sensing applications. *RSC Adv.* **6**, 21624–21661 (2016).
15. Wu, D. *et al.* Fluorescent chemosensors: the past, present and future. *Chem. Soc. Rev.* **46**, 7105–7123 (2017).
16. Yin, H., Liao, L. & Fang, J. Enhanced Permeability and Retention (EPR) Effect Based Tumor Targeting: The Concept, Application and Prospect. 5 (2014).

17. Yu, M. & Zheng, J. Clearance Pathways and Tumor Targeting of Imaging Nanoparticles. *ACS Nano* **9**, 6655–6674 (2015).
18. Kircher, M. F. *et al.* A brain tumor molecular imaging strategy using a new triple-modality MRI-photoacoustic-Raman nanoparticle. *Nat. Med.* **18**, 829 (2012).
19. Qian, X. *et al.* In vivo tumor targeting and spectroscopic detection with surface-enhanced Raman nanoparticle tags. *Nat. Biotechnol.* **26**, 83 (2008).
20. Bruchez, M., Moronne, M., Gin, P., Weiss, S. & Alivisatos, A. P. Semiconductor Nanocrystals as Fluorescent Biological Labels. *Science* **281**, 2013–2016 (1998).
21. Reisch, A. & Klymchenko, A. S. Fluorescent Polymer Nanoparticles Based on Dyes: Seeking Brighter Tools for Bioimaging. *Small* **12**, 1968–1992 (2016).
22. Thiyagarajan, S. K., Raghupathy, S., Palanivel, D., Raji, K. & Ramamurthy, P. Fluorescent carbon nano dots from lignite: unveiling the impeccable evidence for quantum confinement. *Phys. Chem. Chem. Phys.* **18**, 12065–12073 (2016).
23. Huang, Z. *et al.* Quantum confinement in graphene quantum dots. *Phys. Status Solidi RRL – Rapid Res. Lett.* **8**, 436–440 (2014).
24. Ding, H., Yu, S.-B., Wei, J.-S. & Xiong, H.-M. Full-Color Light-Emitting Carbon Dots with a Surface-State-Controlled Luminescence Mechanism. *ACS Nano* **10**, 484–491 (2016).
25. Dhenadhayalan, N., Lin, K.-C., Suresh, R. & Ramamurthy, P. Unravelling the Multiple Emissive States in Citric-Acid-Derived Carbon Dots. *J. Phys. Chem. C* **120**, 1252–1261 (2016).
26. De Luca, M. *et al.* Advances in Use of Capsule-Based Fluorescent Sensors for Measuring Acidification of Endocytic Compartments in Cells with Altered Expression of V-ATPase Subunit V1G1. *ACS Appl. Mater. Interfaces* **7**, 15052–15060 (2015).
27. Acosta, M. A., Ymele-Leki, P., Kostov, Y. V. & Leach, J. B. Fluorescent microparticles for sensing cell microenvironment oxygen levels within 3D scaffolds. *Biomaterials* **30**, 3068–3074 (2009).
28. Stich, M. I. J., Fischer, L. H. & Wolfbeis, O. S. Multiple fluorescent chemical sensing and imaging. *Chem. Soc. Rev.* **39**, 3102–3114 (2010).
29. Carregal-Romero, S. *et al.* Multiplexed Sensing and Imaging with Colloidal Nano- and Microparticles. *Annu. Rev. Anal. Chem.* **6**, 53–81 (2013).
30. del Mercato, L. L., Abbasi, A. Z., Ochs, M. & Parak, W. J. Multiplexed Sensing of Ions with Barcoded Polyelectrolyte Capsules. *ACS Nano* **5**, 9668–9674 (2011).
31. Mercato, L. L. del, Moffa, M., Rinaldi, R. & Pisignano, D. Ratiometric Organic Fibers for Localized and Reversible Ion Sensing with Micrometer-Scale Spatial Resolution. *Small* **11**, 6417–6424 (2015).
32. Leonhardt, H., Gordon, L. & Livingston, R. Acid-base equilibriums of fluorescein and 2',7'-dichlorofluorescein in their ground and fluorescent states. *J. Phys. Chem.* **75**, 245–249 (1971).

33. Ha, T. *et al.* Probing the interaction between two single molecules: fluorescence resonance energy transfer between a single donor and a single acceptor. *Proc. Natl. Acad. Sci.* **93**, 6264–6268 (1996).
34. Lakowicz, J. R. Energy Transfer. in *Principles of Fluorescence Spectroscopy* (ed. Lakowicz, J. R.) 367–394 (Springer US, 1999). doi:10.1007/978-1-4757-3061-6_13
35. Pompa, P. P. *et al.* Fluorescence resonance energy transfer induced by conjugation of metalloproteins to nanoparticles. *Chem. Phys. Lett.* **417**, 351–357 (2006).
36. Chen, G., Song, F., Xiong, X. & Peng, X. Fluorescent Nanosensors Based on Fluorescence Resonance Energy Transfer (FRET). *Ind. Eng. Chem. Res.* **52**, 11228–11245 (2013).
37. Halter, O. & Plenio, H. Fluorescence resonance energy transfer (FRET) for the verification of dual gold catalysis. *Chem. Commun.* **53**, 12461–12464 (2017).
38. Zadran, S. *et al.* Fluorescence resonance energy transfer (FRET)-based biosensors: visualizing cellular dynamics and bioenergetics. *Appl. Microbiol. Biotechnol.* **96**, 895–902 (2012).
39. Stryer, L. & Haugland, R. P. Energy transfer: a spectroscopic ruler. *Proc. Natl. Acad. Sci.* **58**, 719–726 (1967).
40. Wang, J. *et al.* Ratiometric ultrasensitive fluorometric detection of ascorbic acid using a dually emitting CdSe@SiO$_2$@CdTe quantum dot hybrid. *Microchim. Acta* **185**, 42 (2018).
41. Diaz-Diestra, D., Thapa, B., Beltran-Huarac, J., Weiner, B. R. & Morell, G. L-cysteine capped ZnS:Mn quantum dots for room-temperature detection of dopamine with high sensitivity and selectivity. *Biosens. Bioelectron.* **87**, 693–700 (2017).
42. Chen, Z., Lu, D., Cai, Z., Dong, C. & Shuang, S. Bovine serum albumin-confined silver nanoclusters as fluorometric probe for detection of biothiols. *Luminescence* **29**, 722–727 (2014).
43. Xu, H. *et al.* Sodium 4-mercaptophenolate capped CdSe/ZnS quantum dots as a fluorescent probe for pH detection in acidic aqueous media. *Luminescence* **33**, 410–416 (2018).
44. Duong, H. & Rhee, J. Use of CdSe/ZnS core-shell quantum dots as energy transfer donors in sensing glucose. *Talanta* **73**, 899–905 (2007).
45. Liu, H. *et al.* Rapid sonochemical synthesis of highly luminescent non-toxic AuNCs and Au@AgNCs and Cu (II) sensing. *Chem. Commun.* **47**, 4237–4239 (2011).
46. Liu, Y., Ai, K., Cheng, X., Huo, L. & Lu, L. Gold-Nanocluster-Based Fluorescent Sensors for Highly Sensitive and Selective Detection of Cyanide in Water. *Adv. Funct. Mater.* **20**, 951–956 (2010).
47. Nandi, I. *et al.* Protein Fibril-Templated Biomimetic Synthesis of Highly Fluorescent Gold Nanoclusters and Their Applications in Cysteine Sensing. *ACS Omega* (2018). doi:10.1021/acsomega.8b01033

48. Chen, T.H. & Tseng, W.L. (Lysozyme Type VI)-Stabilized Au8 Clusters: Synthesis Mechanism and Application for Sensing of Glutathione in a Single Drop of Blood. *Small* **8**, 1912–1919 (2012).
49. Wen, F. *et al.* Horseradish Peroxidase Functionalized Fluorescent Gold Nanoclusters for Hydrogen Peroxide Sensing. *Anal. Chem.* **83**, 1193–1196 (2011).
50. Shang, L. & Dong, S. Silver nanocluster-based fluorescent sensors for sensitive detection of Cu(II). *J. Mater. Chem.* **18**, 4636–4640 (2008).
51. Enkin, N., Wang, F., Sharon, E., Albada, H. B. & Willner, I. Multiplexed Analysis of Genes Using Nucleic Acid-Stabilized Silver-Nanocluster Quantum Dots. *ACS Nano* **8**, 11666–11673 (2014).
52. Yu, C., Zeng, F., Luo, M. & Wu, S. A silica nanoparticle-based sensor for selective fluorescent detection of homocysteine via interaction differences between thiols and particle-surface-bound polymers. *Nanotechnology* **23**, 305503 (2012).
53. Kim, Y. *et al.* Quantitative analysis of hydrogen peroxide using ratiometric fluorescent probe-doped silica nanoparticles. *Toxicol. Environ. Health Sci.* **9**, 108–115 (2017).
54. Chu, B. *et al.* Fluorescent and Photostable Silicon Nanoparticles Sensors for Real-Time and Long-Term Intracellular pH Measurement in Live Cells. *Anal. Chem.* **88**, 9235–9242 (2016).
55. Shangguan, J. *et al.* Label-Free Carbon-Dots-Based Ratiometric Fluorescence pH Nanoprobes for Intracellular pH Sensing. *Anal. Chem.* **88**, 7837–7843 (2016).
56. Shao, T., Zhang, P., Tang, L., Zhuo, S. & Zhu, C. Highly sensitive enzymatic determination of urea based on the pH-dependence of the fluorescence of graphene quantum dots. *Microchim. Acta* **182**, 1431–1437 (2015).
57. Li, G. *et al.* Facile and Sensitive Fluorescence Sensing of Alkaline Phosphatase Activity with Photoluminescent Carbon Dots Based on Inner Filter Effect. *Anal. Chem.* **88**, 2720–2726 (2016).
58. Shen, P. & Xia, Y. Synthesis-Modification Integration: One-Step Fabrication of Boronic Acid Functionalized Carbon Dots for Fluorescent Blood Sugar Sensing. *Anal. Chem.* **86**, 5323–5329 (2014).
59. Yu, C., Li, X., Zeng, F., Zheng, F. & Wu, S. Carbon-dot-based ratiometric fluorescent sensor for detecting hydrogen sulfide in aqueous media and inside live cells. *Chem. Commun.* **49**, 403–405 (2012).
60. Chandra, A. & Singh, N. Biocompatible Fluorescent Carbon Dots for Ratiometric Intracellular pH Sensing. *ChemistrySelect* **2**, 5723–5728 (2017).
61. Chen, J. *et al.* Novel fluorescent polymeric nanoparticles for highly selective recognition of copper ion and sulfide anion in water. *Sens. Actuators B Chem.* **206**, 230–238 (2015).
62. Wang, L., Zhuo, S., Tang, H. & Cao, D. A near-infrared turn on fluorescent probe for cysteine based on organic nanoparticles. *Sens. Actuators B Chem.* **277**, 437–444 (2018).

63. Melnychuk, N. & Klymchenko, A. S. DNA-Functionalized Dye-Loaded Polymeric Nanoparticles: Ultrabright FRET Platform for Amplified Detection of Nucleic Acids. *J. Am. Chem. Soc.* (2018). doi:10.1021/jacs.8b05840
64. Zhao, Y. *et al.* One-step synthesis of fluorescent organic nanoparticles: The application to label-free ratiometric fluorescent pH sensor. *Sens. Actuators B Chem.* **273**, 1479–1486 (2018).
65. Alivisatos, A. P., Gu, W. & Larabell, C. Quantum Dots as Cellular Probes. *Annu. Rev. Biomed. Eng.* **7**, 55–76 (2005).
66. SMITH, A. M. & NIE, S. Semiconductor Nanocrystals: Structure, Properties, and Band Gap Engineering. *Acc. Chem. Res.* **43**, 190–200 (2010).
67. Zhang, F., Ali, Z., Amin, F., Riedinger, A. & Parak, W. J. In vitro and intracellular sensing by using the photoluminescence of quantum dots. *Anal. Bioanal. Chem.* **397**, 935–942 (2010).
68. Zorman, B., Ramakrishna, M. V. & Friesner, R. A. Quantum Confinement Effects in CdSe Quantum Dots. *J. Phys. Chem.* **99**, 7649–7653 (1995).
69. Samokhvalov, P., Linkov, P., Michel, J., Molinari, M. & Nabiev, I. Photoluminescence quantum yield of CdSe-ZnS/CdS/ZnS core-multishell quantum dots approaches 100% due to enhancement of charge carrier confinement. in (eds. Parak, W. J., Osinski, M. & Yamamoto, K. I.) 89550S (2014). doi:10.1117/12.2040196
70. Hines, M. A. & Guyot-Sionnest, P. Synthesis and Characterization of Strongly Luminescing ZnS-Capped CdSe Nanocrystals. *J. Phys. Chem.* **100**, 468–471 (1996).
71. Chen, N. *et al.* The cytotoxicity of cadmium-based quantum dots. *Biomaterials* **33**, 1238–1244 (2012).
72. Pellegrino, T. *et al.* Hydrophobic Nanocrystals Coated with an Amphiphilic Polymer Shell: A General Route to Water Soluble Nanocrystals. *Nano Lett.* **4**, 703–707 (2004).
73. Fernández-Argüelles, M. T. *et al.* Synthesis and Characterization of Polymer-Coated Quantum Dots with Integrated Acceptor Dyes as FRET-Based Nanoprobes. *Nano Lett.* **7**, 2613–2617 (2007).
74. Ukeda, H., Fujita, Y., Ohira, M. & Sawamura, M. Immobilized Enzyme-Based Microtiter Plate Assay for Glucose in Foods. *J. Agric. Food Chem.* **44**, 3858–3863 (1996).
75. Wu, X. J. & Choi, M. M. F. An optical glucose biosensor based on entrapped-glucose oxidase in silicate xerogel hybridised with hydroxyethyl carboxymethyl cellulose. *Anal. Chim. Acta* **514**, 219–226 (2004).
76. Lepore, M. *et al.* Glucose concentration determination by means of fluorescence emission spectra of soluble and insoluble glucose oxidase: some useful indications for optical fibre-based sensors. *J. Mol. Catal. B Enzym.* **31**, 151–158 (2004).

77. Duong, H. & Rhee, J. Preparation and characterization of sensing membranes for the detection of glucose, lactate and tyramine in microtiter plates. *Talanta* **72**, 1275–1282 (2007).
78. Potter, M., Newport, E. & Morten, K. J. The Warburg effect: 80 years on. *Biochem. Soc. Trans.* **44**, 1499–1505 (2016).
79. Fong, J. F. Y., Chin, S. F. & Ng, S. M. A unique "turn-on" fluorescence signalling strategy for highly specific detection of ascorbic acid using carbon dots as sensing probe. *Biosens. Bioelectron.* **85**, 844–852 (2016).
80. Liu, R. *et al.* Synthesis of glycine-functionalized graphene quantum dots as highly sensitive and selective fluorescent sensor of ascorbic acid in human serum. *Sens. Actuators B Chem.* **241**, 644–651 (2017).
81. Su, H., Sun, B., Chen, L., Xu, Z. & Ai, S. Colorimetric sensing of dopamine based on the aggregation of gold nanoparticles induced by copper ions. *Anal. Methods* **4**, 3981–3986 (2012).
82. Li, L., Liu, H., Shen, Y., Zhang, J. & Zhu, J.-J. Electrogenerated Chemiluminescence of Au Nanoclusters for the Detection of Dopamine. *Anal. Chem.* **83**, 661–665 (2011).
83. Naccarato, A., Gionfriddo, E., Sindona, G. & Tagarelli, A. Development of a simple and rapid solid phase microextraction-gas chromatography–triple quadrupole mass spectrometry method for the analysis of dopamine, serotonin and norepinephrine in human urine. *Anal. Chim. Acta* **810**, 17–24 (2014).
84. Ye, X., Kuklenyik, Z., Needham, L. L. & Calafat, A. M. Automated On-Line Column-Switching HPLC-MS/MS Method with Peak Focusing for the Determination of Nine Environmental Phenols in Urine. *Anal. Chem.* **77**, 5407–5413 (2005).
85. Alarcón-Angeles, G. *et al.* Enhanced host–guest electrochemical recognition of dopamine using cyclodextrin in the presence of carbon nanotubes. *Carbon* **46**, 898–906 (2008).
86. Ma, W. *et al.* Ubiquinone-quantum dot bioconjugates for in vitro and intracellular complex I sensing. *Sci. Rep.* **3**, 1537 (2013).
87. Rakovich, T. Y. *et al.* Highly Sensitive Single Domain Antibody–Quantum Dot Conjugates for Detection of HER2 Biomarker in Lung and Breast Cancer Cells. *ACS Nano* **8**, 5682–5695 (2014).
88. Jeong, S. *et al.* Cancer-Microenvironment-Sensitive Activatable Quantum Dot Probe in the Second Near-Infrared Window. *Nano Lett.* **17**, 1378–1386 (2017).
89. Medintz, I. L. *et al.* Quantum-dot/dopamine bioconjugates function as redox coupled assemblies for *in vitro* and intracellular pH sensing. *Nat. Mater.* **9**, 676 (2010).
90. Lee, K. R. & Kang, I.-J. Effects of dopamine concentration on energy transfer between dendrimer–QD and dye-labeled antibody. *Ultramicroscopy* **109**, 894–898 (2009).

91. Amelia, M., Lavie-Cambot, A., McClenaghan, N. D. & Credi, A. A ratiometric luminescent oxygen sensor based on a chemically functionalized quantum dot. *Chem. Commun.* **47**, 325–327 (2010).
92. Kong, C. *et al.* Determination of dissolved oxygen based on photoinduced electron transfer from quantum dots to methyl viologen. *Anal. Methods* **2**, 1056–1062 (2010).
93. Lemon, C. M. *et al.* Micelle-Encapsulated Quantum Dot-Porphyrin Assemblies as *in Vivo* Two-Photon Oxygen Sensors. *J. Am. Chem. Soc.* **137**, 9832–9842 (2015).
94. Shamirian, A., Samareh Afsari, H., Hassan, A., Miller, L. W. & Snee, P. T. In Vitro Detection of Hypoxia Using a Ratiometric Quantum Dot-Based Oxygen Sensor. *ACS Sens.* **1**, 1244–1250 (2016).
95. Jimenez de Aberasturi, D. *et al.* Optical Sensing of Small Ions with Colloidal Nanoparticles. *Chem. Mater.* **24**, 738–745 (2012).
96. Zeng, S. *et al.* A Review on Functionalized Gold Nanoparticles for Biosensing Applications. *Plasmonics* **6**, 491 (2011).
97. Saha, K., Agasti, S. S., Kim, C., Li, X. & Rotello, V. M. Gold Nanoparticles in Chemical and Biological Sensing. *Chem. Rev.* **112**, 2739–2779 (2012).
98. Aldewachi, H. *et al.* Gold nanoparticle-based colorimetric biosensors. *Nanoscale* **10**, 18–33 (2017).
99. Riedinger, A. *et al.* Ratiometric Optical Sensing of Chloride Ions with Organic Fluorophore–Gold Nanoparticle Hybrids: A Systematic Study of Design Parameters and Surface Charge Effects. *Small* **6**, 2590–2597 (2010).
100. Amin, F., Yushchenko, D. A., Montenegro, J. M. & Parak, W. J. Integration of Organic Fluorophores in the Surface of Polymer-Coated Colloidal Nanoparticles for Sensing the Local Polarity of the Environment. *ChemPhysChem* **13**, 1030–1035 (2012).
101. Li, H. *et al.* Silver Nanoparticle-Enhanced Fluorescence Resonance Energy Transfer Sensor for Human Platelet-Derived Growth Factor-BB Detection. *Anal. Chem.* **85**, 4492–4499 (2013).
102. Song, L., Hennink, E. J., Young, I. T. & Tanke, H. J. Photobleaching kinetics of fluorescein in quantitative fluorescence microscopy. *Biophys. J.* **68**, 2588–2600 (1995).
103. Widengren, J., Chmyrov, A., Eggeling, C., Löfdahl, P.-Å. & Seidel, C. A. M. Strategies to Improve Photostabilities in Ultrasensitive Fluorescence Spectroscopy. *J. Phys. Chem. A* **111**, 429–440 (2007).
104. Ou, G. *et al.* Fabrication and application of noble metal nanoclusters as optical sensors for toxic metal ions. *Anal. Bioanal. Chem.* **410**, 2485–2498 (2018).
105. Zhang, L. & Wang, E. Metal nanoclusters: New fluorescent probes for sensors and bioimaging. *Nano Today* **9**, 132–157 (2014).

106. Li, J., Zhu, J.-J. & Xu, K. Fluorescent metal nanoclusters: From synthesis to applications. *TrAC Trends Anal. Chem.* **58**, 90–98 (2014).
107. Zhang, M. *et al.* A Highly Selective Fluorescence Turn-on Sensor for Cysteine/Homocysteine and Its Application in Bioimaging. *J. Am. Chem. Soc.* **129**, 10322–10323 (2007).
108. Ali, F. *et al.* A fluorescent probe for specific detection of cysteine in the lipid dense region of cells. *Chem. Commun.* **51**, 16932–16935 (2015).
109. Govindaraju, S., Ankireddy, S. R., Viswanath, B., Kim, J. & Yun, K. Fluorescent Gold Nanoclusters for Selective Detection of Dopamine in Cerebrospinal fluid. *Sci. Rep.* **7**, 40298 (2017).
110. Chen, Z., Zhang, C., Zhou, T. & Ma, H. Gold nanoparticle based colorimetric probe for dopamine detection based on the interaction between dopamine and melamine. *Microchim. Acta* **182**, 1003–1008 (2015).
111. Hou, J., Xu, C., Zhao, D. & Zhou, J. Facile fabrication of hierarchical nanoporous AuAg alloy and its highly sensitive detection towards dopamine and uric acid. *Sens. Actuators B Chem.* **225**, 241–248 (2016).
112. Zhu, L. *et al.* Highly sensitive determination of dopamine by a turn-on fluorescent biosensor based on aptamer labeled carbon dots and nano-graphite. *Sens. Actuators B Chem.* **231**, 506–512 (2016).
113. Xie, J., Zheng, Y. & Ying, J. Y. Protein-Directed Synthesis of Highly Fluorescent Gold Nanoclusters. *J. Am. Chem. Soc.* **131**, 888–889 (2009).
114. Lin, Y.-H. & Tseng, W.-L. Ultrasensitive Sensing of Hg2+ and CH3Hg+ Based on the Fluorescence Quenching of Lysozyme Type VI-Stabilized Gold Nanoclusters. *Anal. Chem.* **82**, 9194–9200 (2010).
115. Liu, C.-L. *et al.* Insulin-Directed Synthesis of Fluorescent Gold Nanoclusters: Preservation of Insulin Bioactivity and Versatility in Cell Imaging. *Angew. Chem. Int. Ed.* **50**, 7056–7060 (2011).
116. Hu, L. *et al.* Highly sensitive fluorescent detection of trypsin based on BSA-stabilized gold nanoclusters. *Biosens. Bioelectron.* **32**, 297–299 (2012).
117. Huang, S., Yao, H., Wang, W., Zhang, J.-R. & Zhu, J.-J. Highly sensitive fluorescence quantification of intracellular telomerase activity by repeat G-rich DNA enhanced silver nanoclusters. *J. Mater. Chem. B* **6**, 4583–4591 (2018).
118. See, W. & Smith, J. URINARY LEVELS OF ACTIVATED TRYPSIN IN WHOLE -ORGAN PANCREAS TRANSPLANT PATIENTS WITH DUODENOCYSTOSTOMIES. *Transplantation* **52**, 630–633 (1991).
119. Xie, J., Zheng, Y. & Ying, J. Y. Highly selective and ultrasensitive detection of Hg(2+) based on fluorescence quenching of Au nanoclusters by Hg(2+)-Au(+) interactions. *Chem. Commun. Camb. Engl.* **46**, 961–963 (2010).

120. Wei, H. *et al.* Lysozyme-stabilized gold fluorescent cluster: Synthesis and application as Hg2+ sensor. *Analyst* **135**, 1406–1410 (2010).
121. Jin, L. *et al.* Biomolecule-stabilized Au nanoclusters as a fluorescence probe for sensitive detection of glucose. *Biosens. Bioelectron.* **26**, 1965–1969 (2011).
122. Lan, G. Y., Chen, W. Y. & Chang, H. T. One-pot synthesis of fluorescent oligonucleotide Ag nanoclusters for specific and sensitive detection of DNA. *Biosens. Bioelectron.* **26**, 2431–2435 (2011).
123. Shang, L., Wang, Y., Huang, L. & Dong, S. Preparation of DNA–Silver Nanohybrids in Multilayer Nanoreactors by in Situ Electrochemical Reduction, Characterization, and Application. *Langmuir* **23**, 7738–7744 (2007).
124. Berti, L. & Burley, G. A. Nucleic acid and nucleotide-mediated synthesis of inorganic nanoparticles. *Nat. Nanotechnol.* **3**, 81–87 (2008).
125. Iyer, K. S., Bond, C. S., Saunders, M. & Raston, C. L. Confinement of Silver Triangles in Silver Nanoplates Templated by Duplex DNA. *Cryst. Growth Des.* **8**, 1451–1453 (2008).
126. Furukawa, S. & Miyasato, T. Three-Dimensional Quantum Well Effects in Ultrafine Silicon Particles. *Jpn. J. Appl. Phys.* **27**, L2207 (1988).
127. Biteen, J. S., Lewis, N. S., Atwater, H. A. & Polman, A. Size-dependent oxygen-related electronic states in silicon nanocrystals. *Appl. Phys. Lett.* **84**, 5389–5391 (2004).
128. Derr, J. *et al.* Quantum confinement regime in silicon nanocrystals. *Phys. E Low-Dimens. Syst. Nanostructures* **41**, 668–670 (2009).
129. Bae, S. W., Tan, W. & Hong, J.-I. Fluorescent dye-doped silica nanoparticles: new tools for bioapplications. *Chem. Commun.* **48**, 2270–2282 (2012).
130. Stöber, W., Fink, A. & Bohn, E. Controlled growth of monodisperse silica spheres in the micron size range. *J. Colloid Interface Sci.* **26**, 62–69 (1968).
131. Bagwe, R. P., Yang, C., Hilliard, L. R. & Tan, W. Optimization of Dye-Doped Silica Nanoparticles Prepared Using a Reverse Microemulsion Method. *Langmuir* **20**, 8336–8342 (2004).
132. Nooney, R. *et al.* Synthesis and characterisation of far-red fluorescent cyanine dye doped silica nanoparticles using a modified microemulsion method for application in bioassays. *Sens. Actuators B Chem.* **221**, 470–479 (2015).
133. Riccò, R. *et al.* Ultra-small dye-doped silica nanoparticles via modified sol-gel technique. *J. Nanoparticle Res.* **20**, 117 (2018).
134. Ma, S. *et al.* One-Step Synthesis of Water-Dispersible and Biocompatible Silicon Nanoparticles for Selective Heparin Sensing and Cell Imaging. *Anal. Chem.* **88**, 10474–10481 (2016).
135. Cai, L. *et al.* Butterfly-Shaped Conjugated Oligoelectrolyte/Graphene Oxide Integrated Assay for Light-Up Visual Detection of Heparin. *Anal. Chem.* **83**, 7849–7855 (2011).

136. Cao, R. & Li, B. A simple and sensitive method for visual detection of heparin using positively-charged gold nanoparticles as colorimetric probes. *Chem. Commun.* **47**, 2865–2867 (2011).
137. Fu, X. *et al.* Label-free colorimetric sensor for ultrasensitive detection of heparin based on color quenching of gold nanorods by graphene oxide. *Biosens. Bioelectron.* **34**, 227–231 (2012).
138. Wang, J. & Qiu, J. A review of carbon dots in biological applications. *J. Mater. Sci.* **51**, 4728–4738 (2016).
139. Zheng, P. & Wu, N. Fluorescence and Sensing Applications of Graphene Oxide and Graphene Quantum Dots: A Review. *Chem. – Asian J.* **12**, 2343–2353 (2017).
140. Lim, S. Y., Shen, W. & Gao, Z. Carbon quantum dots and their applications. *Chem. Soc. Rev.* **44**, 362–381 (2014).
141. Arcudi, F., Đorđević, L. & Prato, M. Synthesis, Separation, and Characterization of Small and Highly Fluorescent Nitrogen-Doped Carbon NanoDots. *Angew. Chem. Int. Ed.* **55**, 2107–2112 (2016).
142. Pan, X. *et al.* Carbon dots originated from methyl red with molecular state and surface state controlled emissions for sensing and imaging. *J. Lumin.* **204**, 303–311 (2018).
143. Miao, P. *et al.* Recent advances in carbon nanodots: synthesis, properties and biomedical applications. *Nanoscale* **7**, 1586–1595 (2015).
144. Guo, L. *et al.* Bottom-up preparation of nitrogen doped carbon quantum dots with green emission under microwave-assisted hydrothermal treatment and their biological imaging. *Mater. Sci. Eng. C* **84**, 60–66 (2018).
145. Yang, Y., Wu, D., Han, S., Hu, P. & Liu, R. Bottom-up fabrication of photoluminescent carbon dots with uniform morphology via a soft–hard template approach. *Chem. Commun.* **49**, 4920–4922 (2013).
146. Li, H. *et al.* One-step ultrasonic synthesis of water-soluble carbon nanoparticles with excellent photoluminescent properties. *Carbon* **49**, 605–609 (2011).
147. Dong, Y., Cai, J., You, X. & Chi, Y. Sensing applications of luminescent carbon based dots. *Analyst* **140**, 7468–7486 (2015).
148. Jelinek, R. Carbon-Dots in Sensing Applications. in *Carbon Quantum Dots: Synthesis, Properties and Applications* (ed. Jelinek, R.) 71–91 (Springer International Publishing, 2017). doi:10.1007/978-3-319-43911-2_6
149. Sun, X. & Lei, Y. Fluorescent carbon dots and their sensing applications. *TrAC Trends Anal. Chem.* **89**, 163–180 (2017).
150. Nie, H. *et al.* Carbon Dots with Continuously Tunable Full-Color Emission and Their Application in Ratiometric pH Sensing. *Chem. Mater.* **26**, 3104–3112 (2014).

151. Qu, K., Wang, J., Ren, J. & Qu, X. Carbon Dots Prepared by Hydrothermal Treatment of Dopamine as an Effective Fluorescent Sensing Platform for the Label-Free Detection of Iron(III) Ions and Dopamine. *Chem. – Eur. J.* **19**, 7243–7249 (2013).
152. Wu, J. *et al.* Detection of lysozyme with aptasensor based on fluorescence resonance energy transfer from carbon dots to graphene oxide. *Luminescence* **31**, 1207–1212 (2016).
153. Zor, E. *et al.* Graphene Quantum Dots-based Photoluminescent Sensor: A Multifunctional Composite for Pesticide Detection. *ACS Appl. Mater. Interfaces* **7**, 20272–20279 (2015).
154. Wang, M. *et al.* "Off–On" fluorescent sensing of organophosphate pesticides using a carbon dot–Au(III) complex. *RSC Adv.* **8**, 11551–11556 (2018).
155. Lin, B. *et al.* Modification-free carbon dots as turn-on fluorescence probe for detection of organophosphorus pesticides. *Food Chem.* **245**, 1176–1182 (2018).
156. Li, H., Yan, X., Lu, G. & Su, X. Carbon dot-based bioplatform for dual colorimetric and fluorometric sensing of organophosphate pesticides. *Sens. Actuators B Chem.* **260**, 563–570 (2018).
157. Gao, X., Du, C., Zhuang, Z. & Chen, W. Carbon quantum dot-based nanoprobes for metal ion detection. *J. Mater. Chem. C* **4**, 6927–6945 (2016).
158. Peng, D., Zhang, L., Liang, R.-P. & Qiu, J.-D. Rapid Detection of Mercury Ions Based on Nitrogen-Doped Graphene Quantum Dots Accelerating Formation of Manganese Porphyrin. *ACS Sens.* **3**, 1040–1047 (2018).
159. Xu, J. *et al.* Flavonoid moiety-incorporated carbon dots for ultrasensitive and highly selective fluorescence detection and removal of Pb2+. *Nano Res.* **11**, 3648–3657 (2018).
160. Wu, W., Zhou, T., Berliner, A., Banerjee, P. & Zhou, S. Glucose-Mediated Assembly of Phenylboronic Acid Modified CdTe/ZnTe/ZnS Quantum Dots for Intracellular Glucose Probing. *Angew. Chem. Int. Ed.* **49**, 6554–6558 (2010).
161. Yang, W. *et al.* MOF-76: from a luminescent probe to highly efficient UVI sorption material. *Chem. Commun.* **49**, 10415–10417 (2013).
162. Chai, L. *et al.* Functionalized Carbon Quantum Dots with Dopamine for Tyrosinase Activity Monitoring and Inhibitor Screening: In Vitro and Intracellular Investigation. *ACS Appl. Mater. Interfaces* **7**, 23564–23574 (2015).
163. Li, C. *et al.* Red fluorescent carbon dots with phenylboronic acid tags for quick detection of Fe(III) in PC12 cells. *J. Colloid Interface Sci.* **526**, 487–496 (2018).
164. Tang, Z., Lin, Z., Li, G. & Hu, Y. Amino Nitrogen Quantum Dots-Based Nanoprobe for Fluorescence Detection and Imaging of Cysteine in Biological Samples. *Anal. Chem.* **89**, 4238–4245 (2017).
165. Bhattacharyya, S., Prashanthi, S., Bangal, P. R. & Patra, A. Photophysics and Dynamics of Dye-Doped Conjugated Polymer Nanoparticles by Time-Resolved and Fluorescence Correlation Spectroscopy. *J. Phys. Chem. C* **117**, 26750–26759 (2013).

166. Wu, X. & Zhu, W. Stability enhancement of fluorophores for lighting up practical application in bioimaging. *Chem. Soc. Rev.* **44**, 4179–4184 (2015).
167. Behnke, T. *et al.* Encapsulation of Hydrophobic Dyes in Polystyrene Micro- and Nanoparticles via Swelling Procedures. *J. Fluoresc.* **21**, 937–944 (2011).
168. Li, C. & Liu, S. Polymeric assemblies and nanoparticles with stimuli-responsive fluorescence emission characteristics. *Chem. Commun.* **48**, 3262–3278 (2012).
169. Rao, J. P. & Geckeler, K. E. Polymer nanoparticles: Preparation techniques and size-control parameters. *Prog. Polym. Sci.* **36**, 887–913 (2011).
170. Qian, L., Ahmed, A. & Zhang, H. Formation of organic nanoparticles by solvent evaporation within porous polymeric materials. *Chem. Commun.* **47**, 10001–10003 (2011).
171. Wang, Y., Li, P., Truong-Dinh Tran, T., Zhang, J. & Kong, L. Manufacturing Techniques and Surface Engineering of Polymer Based Nanoparticles for Targeted Drug Delivery to Cancer. *Nanomaterials* **6**, 26 (2016).
172. Hornig, S., Heinze, T., Becer, C. R. & Schubert, U. S. Synthetic polymeric nanoparticles by nanoprecipitation. *J. Mater. Chem.* **19**, 3838–3840 (2009).
173. Liu, M. *et al.* Formation of poly(l,d-lactide) spheres with controlled size by direct dialysis. *Polymer* **48**, 5767–5779 (2007).
174. Byrappa, K., Ohara, S. & Adschiri, T. Nanoparticles synthesis using supercritical fluid technology – towards biomedical applications. *Adv. Drug Deliv. Rev.* **60**, 299–327 (2008).
175. Leobandung, W., Ichikawa, H., Fukumori, Y. & Peppas, N. A. Monodisperse nanoparticles of poly(ethylene glycol) macromers and N-isopropyl acrylamide for biomedical applications. *J. Appl. Polym. Sci.* **87**, 1678–1684 (2003).
176. Li, W.-H. & Stöver, H. D. H. Monodisperse Cross-Linked Core−Shell Polymer Microspheres by Precipitation Polymerization. *Macromolecules* **33**, 4354–4360 (2000).
177. Li, X.-G., Zhang, J.-L. & Huang, M.-R. Interfacial Synthesis and Functionality of Self-Stabilized Polydiaminonaphthalene Nanoparticles. *Chem. – Eur. J.* **18**, 9877–9885 (2012).
178. Monteiro, M. J. & Cunningham, M. F. Polymer Nanoparticles via Living Radical Polymerization in Aqueous Dispersions: Design and Applications. *Macromolecules* **45**, 4939–4957 (2012).
179. Wang, X., Liu, L., Zhu, S. & Li, L. Preparation of organic fluorescent nanocomposites and their application in DNA detection. *Colloids Surf. Physicochem. Eng. Asp.* **520**, 72–77 (2017).
180. Choi, H. M. T. *et al.* Programmable in situ amplification for multiplexed imaging of mRNA expression. *Nat. Biotechnol.* **28**, 1208–1212 (2010).
181. Ali, M. M. *et al.* Rolling circle amplification: a versatile tool for chemical biology, materials science and medicine. *Chem. Soc. Rev.* **43**, 3324–3341 (2014).
182. Huang, J. *et al.* Fluorescence resonance energy transfer-based hybridization chain reaction for in situ visualization of tumor-related mRNA. *Chem. Sci.* **7**, 3829–3835 (2016).

183. Wang, S., Gaylord, B. S. & Bazan, G. C. Fluorescein Provides a Resonance Gate for FRET from Conjugated Polymers to DNA Intercalated Dyes. *J. Am. Chem. Soc.* **126**, 5446–5451 (2004).
184. Dmitriev, R. I. *et al.* Versatile Conjugated Polymer Nanoparticles for High-Resolution O2 Imaging in Cells and 3D Tissue Models. *ACS Nano* **9**, 5275–5288 (2015).